\documentclass[twocolumn,prl,aps,twocolumn]{revtex4-1}
\usepackage[latin9]{inputenc}
\usepackage{amsbsy}
\usepackage{relsize}
\usepackage{amstext}
\usepackage{soul}
\usepackage{bbm}
\usepackage[unicode=true,pdfusetitle,
 bookmarks=false,
 breaklinks=false,pdfborder={0 0 1},backref=false,colorlinks=false]
 {hyperref}
\hypersetup{
 bookmarksnumbered=false,bookmarksopen=false}

\newcommand{\id}{\mathbbm{1}}

\makeatletter

\@ifundefined{textcolor}{}
{%
 \definecolor{BLACK}{gray}{0}
 \definecolor{WHITE}{gray}{1}
 \definecolor{RED}{rgb}{1,0,0}
 \definecolor{GREEN}{rgb}{0,1,0}
 \definecolor{BLUE}{rgb}{0,0,1}
 \definecolor{CYAN}{cmyk}{1,0,0,0}
 \definecolor{MAGENTA}{cmyk}{0,1,0,0}
 \definecolor{YELLOW}{cmyk}{0,0,1,0}
}

\usepackage{amsmath}
\usepackage{graphicx}
\usepackage{amssymb}
\usepackage{txfonts,color}

\makeatother

\begin{document}

\title{Noise-resilient Quantum Computing with a Nitrogen-Vacancy Center and Nuclear Spins}
\author{J. Casanova}
\affiliation{Institut f\"ur Theoretische Physik and IQST, Albert-Einstein-Allee 11, Universit\"at
Ulm, D-89069 Ulm, Germany}
\author{Z.-Y. Wang}
\affiliation{Institut f\"ur Theoretische Physik and IQST, Albert-Einstein-Allee 11, Universit\"at
Ulm, D-89069 Ulm, Germany}
\author{M. B. Plenio}
\affiliation{Institut f\"ur Theoretische Physik and IQST, Albert-Einstein-Allee 11, Universit\"at
Ulm, D-89069 Ulm, Germany}

\begin{abstract}
Selective control of qubits in a quantum register for the purposes of quantum information
processing represents a critical challenge for dense spin ensembles in solid state systems. Here we present a protocol that achieves a complete set of selective electron-nuclear gates and single nuclear rotations in such an ensemble in diamond facilitated by a nearby 
NV center. The protocol suppresses internuclear interactions as well as unwanted coupling 
between the NV center and other spins of the ensemble to achieve quantum gate fidelities well 
exceeding 99\%. Notably, our method can be applied to weakly coupled, distant, spins  representing a scalable 
procedure that exploits the exceptional properties of nuclear spins in diamond as robust quantum 
memories.
\end{abstract}
\maketitle

{\em Introduction --} Quantum computing and quantum simulation hold the promise for tackling computational 
problems that are currently out of the reach of classical devices~\cite{Feynman82, Lloyd96}. 
With these applications in mind a wide variety of possible platforms have been proposed 
and realized which include trapped ions~\cite{LeibfriedEtAl}, superconducting circuits~\cite{Devoret13}, 
optical lattices~\cite{Bloch05}, coupled cavity arrays \cite{HartmannBP08}, integrated 
photonics \cite{Obrien09}, and hybrid systems involving nuclear spins and nitrogen vacancy (NV) 
centers \cite{Doherty13}. However, while the storage and processing of information by using quantum
degrees of freedom promises to enhance our computational capabilities, the available quantum-bits (qubits) are  fragile and strongly sensitive to environmental fluctuations.

Nuclear spin clusters in materials such as diamond have been identified as promising candidates
for robust solid-state quantum memories because of their long coherence times and the potentially
large number of available spins \cite{Doherty13}. Nuclear spins can be initialized, controlled,
and read out for quantum information processing and sensing purposes with an NV center driven by optical fields and  microwave radiation \cite{GruberDT+1997,
GaebelDP+2006,Gurudev07,Neuman10, Robledo11, vanderSar12, Kolkowitz12, Taminiau12, Liu13,Taminiau14, WaldherrWZ+2014}. However quantum computing requires the precise manipulation of the information encoded in each  qubit which becomes 
a delicate issue in samples with dense resonance spectra. Additionally, while nuclear spins can be effectively isolated from other spins~\cite{Maurer12}, to combine this protection with a sequential generation of a complete set of quantum gates on specific nuclei remains as a challenging task. Overcoming these issues would enable us to realize circuit-based algorithms in highly polarized nuclear  registers \cite{Nielsen} as
well as alternative models such as DQC1 computing that do not require initial nuclear polarization
\cite{KnillL1998,ParkerP2000}.

In this Letter we present a protocol that combines the advantages of recently developed
dynamical decoupling protocols \cite{Casanova15,Wang15} for the suppression of both electronic 
decoherence and internuclear interactions with the robust and selective implementation of quantum gates. Following a description of the technical details of our method that incorporates the combined action of microwave and radio frequency fields, we show the existence of a low-energy branch that is useful for individual qubit coherent control. Finally we proceed to demonstrate with detailed numerical simulations that our scheme allows for protected single-qubit rotations and two-qubit gates between an NV center and weakly coupled $^{13}$C-nuclei and can achieve fidelities above 99\%~\cite{Fowler12}. Additionally, and although we use NV centers on diamond as the model system, our method is equally applicable to other platforms as the case of phosphorus in silicon, or silicon carbide.

{\em Formalism --} Let us consider an NV center  in a nuclear spin bath where a static
magnetic field $B_z$  is applied along the NV axis (the $\hat{z}$ axis). Microwave and rf fields
are used for external control over the electron and nuclear spins as well as for achieving internuclear
decoupling. The Hamiltonian that describes this situation reads ($\hbar = 1$)
\begin{equation}\label{model}
    H = DS_{z}^{2} - \gamma_{e}B_{z}S_{z} - \sum_{j} \gamma_{n} B_{z} \ I_{j}^{z} +
    S_{z}\sum_{j}\vec{A}_{j}\cdot\vec{I}_j + H_{\rm nn} + H_{\rm c}.
\end{equation}
Here $\gamma_{e}$ ($\gamma_{n}$) is the electronic (nuclear) gyromagnetic ratio,  $H_c$ describes the action of microwave and rf control fields (detailed below),
and $H_{\rm nn}$ accounts for the internuclear coupling \cite{Supplemental}. The interaction between
the NV center and the $j$-th nucleus is mediated by the hyperfine vector $\vec{A}_{j}$. Note that
due to the large zero field splitting $D=2\pi\times2.87$ GHz, we have eliminated non-secular components
in Eq.~(\ref{model}). The microwave field is tuned to address specific nuclear spins thanks to
a tailored sequence of $\pi$-pulses, the AXY-$8$ sequence~\cite{Casanova15,Wang15}.
The rf-field contains the decoupling field $\vec{B}_{\rm d} = B_{\rm d} \cos{(\omega_{\rm rf} t)} \ \hat{n}$
and the control field $\vec{B}_{\rm c} = B_{\rm c} \cos{(\omega_{\rm c} t + \phi_c)}  \ \hat{n}_c$.
Additionally and because of recently developed $3$D positioning methods~\cite{Wang15} we can
assume  $\vec{A}_j$, $\hat{n}$, and $\hat{n}_c$ to be known quantities.

The decoupling field $\vec{B}_{\rm d}$ gives rise to the appearance of different branches 
of resonance frequencies of the nuclear spins that can be exploited both, to obtain entangling interactions 
between the electron spin and some specific nuclei as well as single qubit operations, see Eq.~$(21)$
of \cite{Supplemental} for a view of the resonances map. These branches
are $\omega_j$, $\omega_{\rm rf} \pm \omega_j$, $3\omega_{\rm rf} \pm \omega_j$, and $\omega_{\rm rf}$ where $j$ labels each nucleus  and
\begin{equation}\label{resonances}
    \omega_j = |\Delta| \sqrt{ \bigg[\delta + \frac{m_s}{2 \Delta}  \tilde{n}_z \ \tilde{n}_x \ A_j^z \bigg]^2
    + \bigg[\frac{\delta}{\sqrt{2}} + \frac{m_s}{2 \Delta}   \ \tilde{n}^2_z \ A_j^z \bigg]^2},
\end{equation}
with $A_j^z = \vec{A}_j\cdot\hat{z}$, $\delta =  \tilde{n}_x + \frac{\Omega_x}{2 \Delta} (1 - \tilde{n}_x^2 - \tilde{n}_z)$,
$\Omega_x = \Omega \ n_x$, $n_x$ being the $x$-component of the decoupling-field vector $\hat{n}$,
$\Omega = \frac{\gamma_n B_d}{2}$, and $m_s = \pm 1$. The quantities $\tilde{n}_x$ and  $\tilde{n}_z$
accounts for the  $x$- and $z$-components of $\tilde{n} = (\frac{\Omega_x}{\tilde\omega}, 0,  -\frac{\omega
+ \omega_{\rm rf}}{\tilde\omega})$, while $\omega = \gamma _n B_z$, $\tilde{\omega}  = \sqrt{(\omega + \omega_{\rm rf})^2 + \Omega_x^2 }$,
and $\Delta = \tilde{\omega} - 2 \omega_{\rm rf}$.  We  consider $\omega_{\rm rf} \gg \max_j
\omega_j$, a condition that can be achieved with the application of moderate magnetic fields which, 
for the parameter regime considered in this work, need to satisfy $B_z > 0.1$ T.

Restricted to the low-energy resonance branch, i.e. to the set of frequencies $\{\omega_j\}$, the
effective Hamiltonian after eliminating fast rotating terms reads~\cite{Supplemental}
\begin{eqnarray}\label{effective}
    H  &=& \frac{m_s}{2} F(t) \ \sigma_z  \sum_j g_j \bigg[I_j^x \cos{(\omega_j t)} -  I_j^y  \sin{(\omega_j t)}  \bigg]\nonumber\\
       &+& \Omega(t)  \sum_j \vec{I}_j \cdot \bigg[ |\vec{\alpha}_j| \ \hat{\alpha}_j \cos{(\omega_j t)} -  |\vec{\beta}_j| \
       \hat{\beta}_j \sin{(\omega_j t)}  \bigg].
\end{eqnarray}
$F(t) = \pm 1$ is the modulation function, $\Omega(t) = 2\lambda\cos{(\omega_c t +\phi_c)}$ and 
$g_j = \big|  A^z_j \ \tilde{n}_z \sin{(\theta_{\tilde{n},  \hat{n}_j})}  \big|$ with $\theta_{\tilde{n},
\hat{n}_j}$ being the angle between the vectors $\tilde{n}$ and $\hat{n}_j$ with $\hat{n}_j = 
\left( \Delta\delta + m_s/2  \tilde{n}_z \ \tilde{n}_x \ A_j^z, 0, \Delta\delta/\sqrt{2} + m_s/2 \ 
\tilde{n}^2_z \ A_j^z \right)/\omega_j.$  The nuclear quantization axes are  $I_j^x = \vec{I}_j \cdot \hat{x}_j, 
\ I_j^y = \vec{I}_j \cdot \hat{y}_j, \ I_j^z = \vec{I}_j \cdot \hat{z}_j,$ where $\vec{x}_j =  (\vec{\gamma}_{3,j} 
- \vec{\gamma}_{3,j}\cdot \hat{n}_j \ \hat{n}_j)$, $\vec{y}_j=(\hat{n}_j \times \vec{\gamma}_{3,j})$, 
$\vec{z}_j=\hat{n}_j $, $\vec{\gamma}_{3,j} = (\vec{A}_j \cdot \hat{z})(\hat{z} \cdot \tilde{n})\tilde{n}$,
and the vectors $\vec{\alpha}_j$,  $\vec{\beta}_j$ depend on the specific orientation of
$\hat{n}_c$ with respect to the basis $\hat{x}_j$,  $\hat{y}_j$,  $\hat{z}_j$~\cite{Supplemental}.

\begin{figure}[t]
\hspace{-0.30 cm}\includegraphics[width=1\columnwidth]{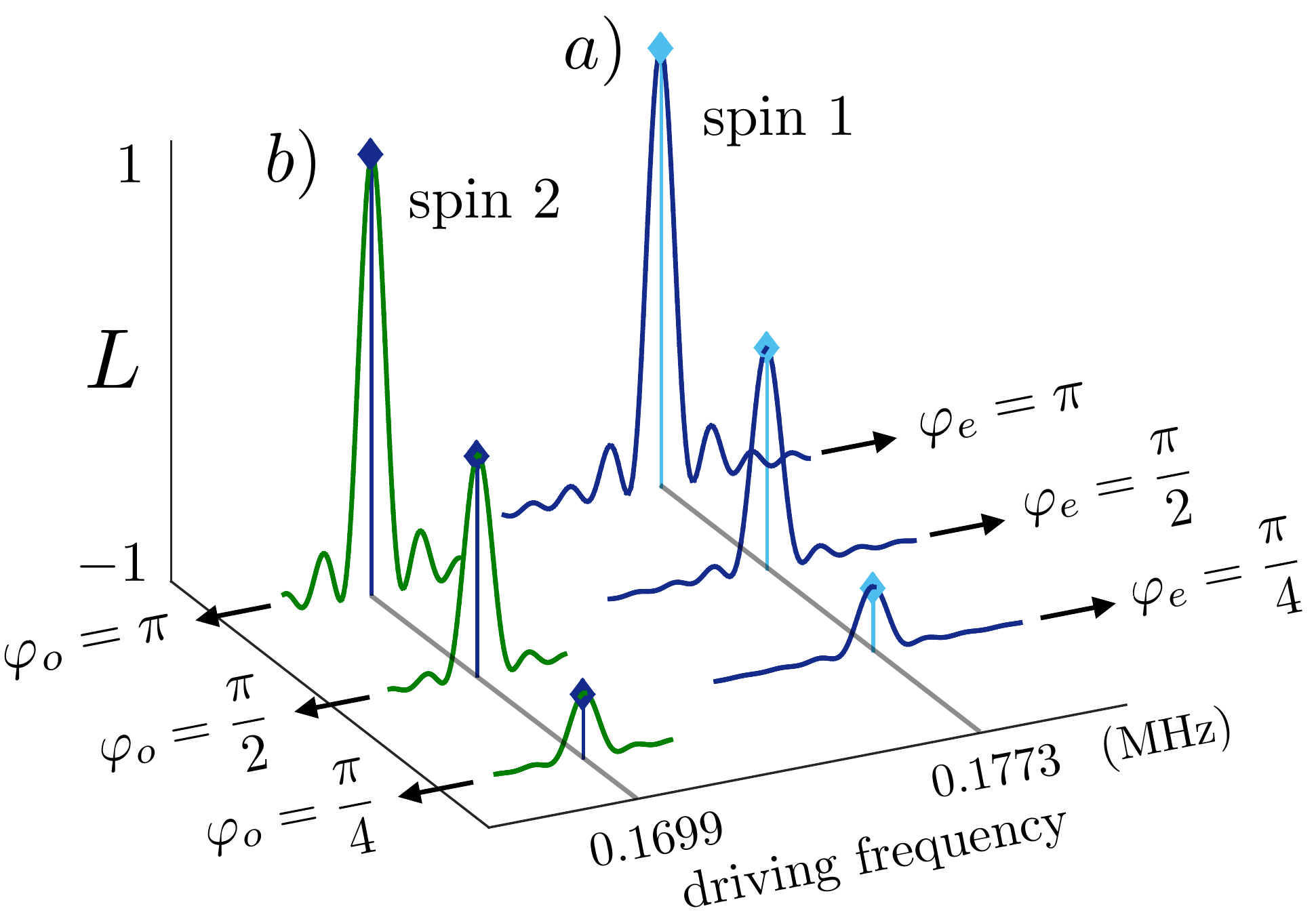}\caption{\label{tunning}(color online)
Coherence ($L$)  evolution as a consequence of the entangling gates mediated by $H^{s}$  a), and  by $H^{a}$ b)
for two different nuclei under the action of a decoupling field $\vec{B}_{\rm d}$ with
$\Delta = 2\pi\times100$ kHz. $\rho \propto |+\rangle\langle +| \otimes I$ is the initial electronic-nuclear
state, with $I$  the identity operator (we assume the nucleus in a thermal state). In all  cases $6000$
imperfect decoupling pulses have been applied  giving rise to an evolution time  of a) $t \approx 3.4$ ms,
and b) $t \approx 3.5$ ms.}
\end{figure}

The contribution of $H_{\rm nn}$ is suppressed in Eq.~(\ref{effective}) 
 allowing to implement quantum operations between the electron spin and different nuclear spins without the 
interference of internuclear interactions. As shown in \cite{Supplemental} this suppression holds if
\begin{eqnarray}
    {\rm max}_j \  | A_j^z| &\ll& |2 \Delta|, \label{f} \\
    \tilde{n}_x + \frac{\Omega_x}{2 \Delta} (1 - \tilde{n}_x^2 - \tilde{n}_z)  &=& \sqrt{2} \
    \big[ \tilde{n}_z + \frac{\Omega_x}{2 \Delta}\tilde{n}_x (1 - \tilde{n}_z) ],\label{s}\\
    \big|\frac{\mu_{0}\gamma^2_{n}}{2|\vec{r}_{j,k}|^{3}}\big|    &\ll& |\omega_j| \label{t},
\end{eqnarray}
where $|\vec{r}_{j,k}|$ is the distance between nuclei $j$ and $k$. Eq.~(\ref{f}) accounts
for the validity of the magic angle condition according to the Lee-Goldburg decoupling
~\cite{Lee65,CaiRJ+2013} for all nuclei and, if satisfied, gives rise to Eq.~(\ref{s}). The
latter corresponds to the magic angle relation in absence of hyperfine fields. Finally Eq.~(\ref{t})
assures that $H_{\rm nn}$ is eliminated up to a factor $ |\frac{A^z_j}{2\Delta} g_{j,k}|$ for each internuclear interaction \cite{Supplemental}. Here $g_{j,k}=\frac{\mu_{0}}{4}
\frac{\gamma_{n}^2}{r_{j,k}^{3}} \left[1 - 3 (n_{j,k}^z)^2 \right]$ are the internuclear coupling coefficients, $ r_{j k}$  the 
distance between the $j$-th and $k$-th nuclei, and $n^z_{j,k}$ the z-component of the unit vector 
$\vec{r}_{j,k}/r_{j,k}$.
In our simulations we consider $\Delta=2\pi\times 100$ kHz which according to Eq.~(\ref{s}) 
result in  $\Omega_x \approx \sqrt{2} \Delta$. Note that  by using external coils rf fields 
of  $\sim 0.1$ T  have been  demonstrated~\cite{Michal08}, and when applied 
to the case of $^{13}$C nuclei  generate $\Omega_x > 2\pi\times 100$ kHz.

{\em Gate performance--} The application of  AXY sequences provides us with a  robust procedure
to entangle the electron spin and each nuclei~\cite{Casanova15}. However, in order to selectively
couple the electron spin with the $I_l^x$ or $I_l^y$ operators we need to choose time-symmetric or antisymmetric pulse sequences, i.e.
a modulation function $F(t)$ as  $F^s(t) = \sum_{k >0} f^{s}_k  \cos{(k\omega t)}$
($F^a(t) = \sum_{k >0} f^{a}_k  \sin{(k\omega t)}$) for the symmetric (anti-symmetric) case~\cite{Supplemental}, where,
importantly for the following, $f^{s}_k$ and $f^{a}_k$ can be continuously tuned~\cite{Casanova15,Wang15}.
In the absence of rf-control, $\lambda = 0$, once the symmetric (anti-symmetric) sequence is applied
and fast rotating terms are eliminated, which requires that the condition  $\left| \frac{f_{\tilde{k}}^{e/o} g_j}{4 (\omega_l - \omega_j)}\right| \ll 1$ $\forall j \neq l$ holds, see Eq.~(\ref{effective})  (note that this is always achievable because of the tunable $f_{\tilde{k}}^{e/o}$) we find $H^{s} = \frac{m_s}{4} f^{s}_{\tilde{k}} g_l
\sigma_z I_l^x$ ($H^{a} = -\frac{m_s}{4} f^{a}_{\tilde{k}} g_l  \sigma_z I_l^y$) as the effective
Hamiltonian from Eq.~(\ref{effective}). In both cases we assume $\tilde{k} \omega = \omega_l$ with
$\tilde{k}\in\mathbb{N}$ and $\omega_l$ the resonance frequency of the $l$-th nucleus. Note that the resonances $\omega_l$ are independent of the magnitude of the $B_z$ field  which allows to work in a wide variety of regimes including the case of high magnetic fields for better elimination  of different rotating terms. When $\vec{B}_c \neq 0$ and $\omega_c = \omega_l$ we can add to $H^{s/a}$ a single-qubit term proportional
to $I_l^x$ or $I_l^y$ by choosing the phase $\phi_c$ in $\Omega(t)$ as $\phi_c = 0$ and $\phi_c =
\frac{\pi}{2}$ respectively. Finally, when the NV center is driven outside of the resonance
band $\{\omega_j\}$ we can eliminate the first line in Eq.~(\ref{effective})  achieving individual
spin rotations. Hence our method provides a universal set of quantum gates.

The time-evolution operators associated to $H^{s/a}$ read $U^{s/a}_t = \exp(-i \frac{m_s}{4}
f^{s/a}_{\tilde{k}} g_l t \  \sigma_z I_l^{x/y} )=\exp(-i \varphi_{s/a}  \sigma_z I_l^{x/y})$. Robustness
of the protocol requires that the time $t$ is a multiple of the period $\tau$ of the pulse sequence,
i.e $t = N \tau$~\cite{Supplemental}. Hence, we are restricted to a set of phases $\varphi_{s/a}\equiv
\varphi_{s/a}(N) = \frac{m_s}{4} f^{s/a}_{\tilde{k}} g_l N \tau$  depending on the number of applied
periods $N$. The absence of tunable coefficients $f^{s/a}_{\tilde{k}}$  would
limit the fidelity of each performed gate. For example the operation  $\exp(-i \frac{\pi}{2}\sigma_z I_l^x )$
requires $\varphi_s(N) = \frac{\pi}{2}$ which, in general, does not hold for standard  sequences
 as CPMG \cite{Carr54, Meiboom58} or the XY family \cite{Maudsley86, Gullion90} where $f^{s/a}_{\tilde{k}}
=\frac{4}{\pi k}$. However, in our case the coefficients $f^{s/a}_{\tilde{k}}$ can be arbitrarily
selected giving access to any value for $\varphi_s(N)$, $\varphi_a(N)$ and to any quantum gate
\cite{Casanova15,Wang15}.

This is shown in Fig.~\ref{tunning} where decoupling pulses are introduced according to  
$H_{\rm mw}=\Omega \cos{[(\omega_{\rm NV} + \Lambda) t}][ \cos{(\phi_i)}\ S_x + \sin{(\phi_i)}\  S_y]$ \cite{Loretz15}, with $\Lambda$  a static detuning error with respect 
to the NV energy transition, $\omega_{\rm NV}$, caused, for example, by a change in temperature  
($\Lambda \approx 2\pi\times 70$ kHz for a 1 K temperature shift~\cite{Faraday15}) or by 
the imperfect polarization of the nitrogen spin of the NV center~\cite{Doherty13} 
($\Lambda \approx 2\pi\times2$ MHz for the $^{14}$N isotope).  The Rabi frequency $\Omega$ is 
chosen to flip the electron spin in $12.5$ ns, and $S_{x,y}$ are spin-1 operators. 
In our numerics we  consider $\vec{A}_{j}=\frac{\mu_{0}\gamma_{e}\gamma_{n}}{2|\vec{r}_{j}|^{3}}[\hat{z} 
- 3\frac{(\hat{z}\cdot\vec{r}_{j})\vec{r}_{j}}{|\vec{r}_{j}|^{2}}]$, with $\vec{r}_j$ connecting the NV center and the nucleus, and the long relaxation time $(T_1)$ of the NV electron 
spin when operated at $T \sim 4$ K. Note that in these conditions $T_1$ measurements on the order of 
many seconds have been reported~\cite{Jarmola12, Cramer15}.

The curves in  a), b) correspond to the evolution of the signal $L = (1-2p_{|\psi_x\rangle} )$ 
with $p_{|\psi_x\rangle} = {\rm Tr}[\rho(t) |\psi_x\rangle\langle\psi_x|]$~\cite{Maze08, Zhao12} 
due to the coherent interaction of the NV center with a nucleus for different values of the driving frequency 
($\tilde{k}\omega$). $|\psi_x\rangle = \sigma_x |\psi_x\rangle$ is an eigenstate of the 
electronic Pauli operator $\sigma_x$. In a) we couple the NV to an spin (spin 1, $A^z = -2\pi\times16.59$ kHz) 
using the symmetric sequence tuned to obtain $\varphi_s = \pi, \frac{\pi}{2}, \frac{\pi}{4}$ 
($f^s_1 = 0.045, 0.0225, 0.0112$) respectively. In b) we use a nucleus (spin 2,  $A^z = 
2\pi\times 9.63$ kHz) and the anti-symmetric sequence with $f^a_1 = 0.071, 0.0355, 0.01775$ 
such that $\varphi_a = \pi, \frac{\pi}{2}, \frac{\pi}{4}$. In all cases we consider 
a detuning error $\Lambda = 2\pi\times 70$ kHz and a Rabi frequency error (RFE) of $0.25 \%$~\cite{Cai12}. 
We note that these results are essentially indistinguishable from the 
case of ideal instantaneous pulses which confirms the robustness of both symmetric  and anti-symmetric 
sequences. The vertical lines are located at the resonance positions predicted by Eq.~(\ref{resonances}) 
and  their height can be calculated by using the expressions for $H^{s/a}$ as $L = -\cos{(\frac{m_s}{4} 
f^{s/a}_{\tilde{k}} g_l   t)}$, for $\frac{m_s}{4} f^{s/a}_{\tilde{k}} g_l   t \equiv \varphi^{s/a} = \pi, 
\frac{\pi}{2}, \frac{\pi}{4}$. Additionally to demonstrate the robustness of the method we 
chose $\tilde{k}= 1$ giving rise to the application of $6000$ imperfect decoupling pulses, however this number 
can be significantly reduced  selecting $\tilde{k} > 1$.

\begin{figure}[t]
\hspace{-0.35 cm}\includegraphics[width=0.9 \columnwidth]{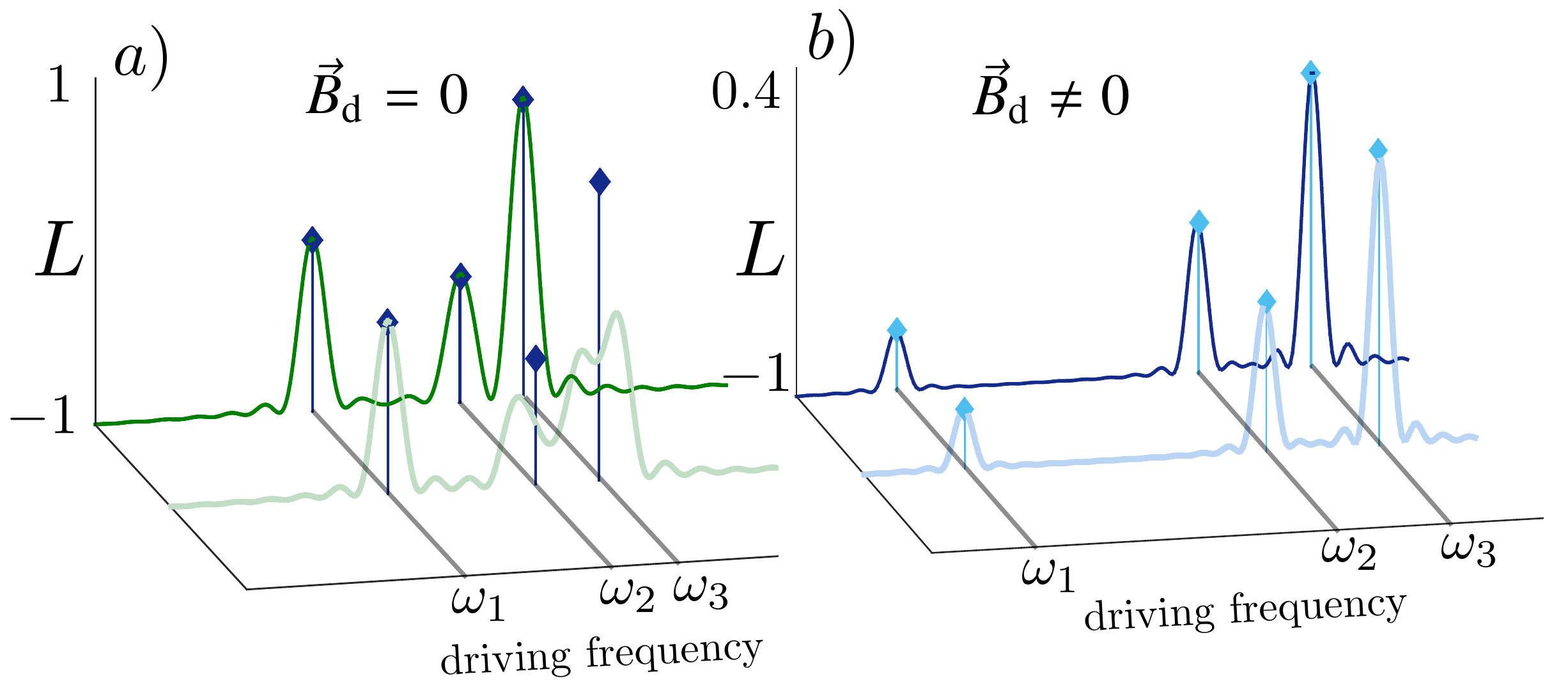}\caption{\label{decoupling}(color 
online) Coherence ($L$) evolution in absence a), and with decoupling field b).  The high internuclear 
coupling distorts the front curve in a) with respect to the case without $H_{\rm nn}$ (back 
curve). In b) we show the results when ${\vec B}_d$ is included giving rise to two overlapping 
figures. The total evolution time is $t \approx 0.81$ ms, a), and $t \approx 3.5$ ms b). The resonance 
frequencies in a) follows the expression $ |\omega \ \hat{z} - \frac{m_s}{2}  \vec{A}_j|$ to obtain 
$\omega_j =2\pi\times (1.0686 ,  1.0770, 1.0745)$ MHz, and in  b) we  use the low-energy brach obtaining 
$\omega_j = 2\pi\times (0.1711,  0.1746,  0.1759)$ MHz.}
\end{figure}

{\em Internuclear decoupling-- } The decoupling efficiency of our method is shown in 
Fig.~\ref{decoupling} where we plot  the evolution of $L$ for a nuclear cluster involving 
a dimer, i.e. a two-qubit register with the nuclei located at the minimum distance 
allowed in the diamond lattice $r_0 \approx 1.54 \  \mathring{\rm A}$, and an isolated 
nucleus. This gives rise to a internuclear coupling coefficient of $2\pi\times0.685$ 
kHz for the dimer, while the coupling of the latter with the additional qubit is smaller 
than  $2\pi\times 1$ Hz. The hyperfine vectors  for each nuclei have $A^z_1 =2\pi\times 4.35$ 
kHz, $A^z_2 = -2\pi\times7.49$ kHz,   $A^z_3 = -2\pi\times11.82$ kHz, and we  apply an 
rf decoupling field with $\Delta = 2\pi\times100$ kHz. $H_{\rm mw}$ accounts for imperfect pulses with $\Lambda = 
2\pi\times70$ kHz, and a RFE of $0.25\%$.

In the absence of the decoupling field ${\vec B}_d$ we can observe the impact of 
the internuclear interactions due to $H_{\rm nn}$. The green curve in the background of 
Fig.~\ref{decoupling} a) ($H_{\rm nn} = 0$) exhibits three clearly identifiable peaks. 
In contrast, the grey curve in foreground of Fig.~\ref{decoupling} a) ($H_{\rm nn}\neq 0$) 
shows the distortion of the resonance peaks of the two nuclei of the dimer 
while the resonance curve of the isolated nucleus remains unchanged. Fig.~\ref{decoupling} 
b) demonstrates the effectiveness of the decoupling field, $\vec{B}_{\rm d}$, as now the 
response in the presence and in the absence of $H_{\rm nn} = 0$ (back dark blue curve and 
front light blue curve) produce the same evolutions for $L$. Again, vertical lines in 
Fig.~\ref{decoupling} a), b) account for the locations of the theoretically predicted
resonances and for their height. For the case  $\vec{B}_{\rm d} = 0$, Fig.~\ref{decoupling} 
a), we have resonances at $ |\omega \ \hat{z} - \frac{m_s}{2}  \vec{A}_j|$, with $\omega = 
\gamma _n B_z$ and $m_s=1$, see for example~\cite{Casanova15}, while the case $\vec{B}_{\rm d} 
\neq 0$, Fig.~\ref{decoupling} b), resonates according to   Eq.~(\ref{resonances}). The 
height of the peaks can be predicted theoretically when dealing with isolated 
nuclear spins, i.e. when $H_{\rm nn} = 0$, or  $H_{\rm nn} \neq 0$ and the decoupling field 
is present. Note that in the  $\vec{B}_{\rm d} = 0$  case we have 
$L = -\cos{(\frac{m_s}{4} f_{\tilde{k}} g_l   t)}$ with $|g_l| = |\vec{A}_l - \vec{A}_l \cdot 
\hat{\omega}_l  \ \hat{\omega}_l|$~\cite{Casanova15}. 

{\em Gate fidelities-- }   To check the fidelities  of single- and two qubit-gates we 
simulate a sample such that $A^z_1 = 2\pi\times 23.10 $ kHz, $A^z_2 = 2\pi\times 9.63$ kHz, 
$A^z_3 = -2\pi\times 16.59$ kHz, the internuclear coupling coefficients are $g_{1, 2} = 2\pi\times 1.64$ Hz, $g_{1, 3} = 2\pi\times 1.40$ Hz, 
$g_{2, 3} = 2\pi\times -186.76$ Hz, and $\Delta =2\pi\times  100$ kHz. This situation involving a minimum distance between the $A^z_j$ components of $\approx 2\pi\times13$ kHz  for, at least, three nuclear spins can be estimated to occur with a $0.06 \%$ of probability in  diamond with $^{13}$C natural abundance. This means that 10 samples of this kind can be found in a diamond layer of dimensions 10 $\mu$m $\times$ 10 $\mu$m $\times$ 10 nm assuming a low NV concentration of 0.01ppm~\cite{Jarmola12}. A relaxation of this condition by looking for an energy difference on the $A^z_j$ components of $2\pi\times 5$ kHz increases the rate of appearance to  $\approx 1 \%$. These estimations have been performed by adding the restriction of discarding samples that present nuclei with $A_j^z > 2\pi\times45$ kHz, which assures a reduction of the internuclear coupling for moderate rf decoupling fields. If one does not include any condition on $A_j^z$, the percentage of suitable samples grows from $0.06\%$ to $5.7\%$. 

We compute the fidelities 
of single- and two- qubit gates according to $F = \frac{|{\rm Tr}(A B^{\dag})|}{\sqrt{{\rm Tr}(A A^{\dag}) 
\ {\rm Tr}(B B^{\dag})}}$, with $A$ and $B$  two general quantum operations~\cite{Wang08}, by 
driving at the theoretically predicted resonance frequencies in Eq.~(\ref{resonances}) with 
imperfect pulses such that $\Lambda = 2\pi\times 70$ kHz and  RFE $=0.25 \%$. The results 
are outlined in Table~\ref{table1} and an inspection confirms the high fidelity achieved by 
our method. See \cite{Supplemental} for further details.

Finally, to estimate the effect of environmental noise from other 
unused nuclei we have completed  the previous sample with a  bath containing $200$ $^{13}$C 
atoms and compare the value of $L$ of after the gate exp$(-i \pi \sigma_zI_1^x)$. This 
situation describes a diamond with a $0.27 \%$ abundance of $^{13}$C  located
within a radius of $\approx 4.7$ nm with the NV assumed to be located in the center.
Because of computational restrictions, here we deal with instantaneous pulses. 
For an initial density matrix  $\rho \propto |+\rangle\langle+| \otimes I_{2^N \times 2^N}$, 
$N$ being the total number of nuclear spins, our realisation of 
exp$(-i \pi \sigma_zI_1^x)$ gives rise to $L =  0.9985$ 
when only the three qubit register is consider and $L =  0.9936$ when the bath is included. 
Note that this value has been obtained by averaging the results for 10 different 
samples~\cite{Supplemental}. This implies that thanks to our decoupling protocols
the large number of bath spins has a small effect on the gate fidelity.

\begin{table}[t]
\centering
\caption{Fidelities $F_{-,+}$ for different nuclear spins in an interacting cluster and 
different entangling- and single-qubit gates. The first number in each box corresponds to 
the fidelity ($F_-$) of the exp$( - i\frac{\pi}{2} \sigma_z I_j^{x, y})$, exp$( - i\frac{\pi}{2}  
I_j^{x, y})$  gates, while the second number to their inverse operations ($F_+$). 
Gate times are $t_1= 480 \ \mu s$ ($800$ pulses), $t_2= 1.6$ ms (2800 pulses), $t_1= 450 \ 
\mu s$ ($800$ pulses).  All the applied pulses are imperfect.}
\label{table1}
\vspace{2.0mm}
\begin{tabular}{{ |c | c | c | c | c| c|}}
 \hline
  $F_{-,+}$                        & exp$( \mp i\frac{\pi}{2} \sigma_z I_j^x)$&  exp$(\mp i\frac{\pi}{2} \sigma_z I_j^y)$ &    exp$( \mp i\frac{\pi}{2}  I_j^x)$ &  exp$( \mp i\frac{\pi}{2}  I_j^y)$\\
\hline
Spin$_1$  & $ 0.9984,  0.9980 $    &$0.9971, 0.9988$      & $0.9983, 0.9983$     &$0.9983, 0.9984$  \\
\hline
Spin$_2$    & $0.9918, 0.9918$    &$0.9935, 0.9930$      & $  0.9986, 0.9986$   & $0.9987, 0.9986$ \\
\hline
Spin$_3$  & $0.9960, 0.9963$    &$0.9975, 0.9963$    & $0.9952, 0.9952$  & $ 0.9954, 0.9953$ \\
\hline
\end{tabular}
\end{table}

{\em Applications--}  An algorithmic problem of  interest in quantum chemistry and solid-state
physics is the quantum simulation of fermionic systems. Through the Jordan-Wigner
transformation~\cite{Jordan28} any fermionic Hamiltonian $H_{f}$ of $N$ particles admits
a form like $H_{f} = \sum_{(i, j, ...)} g_{(i, j, ...)} [\sigma^{\alpha}_{i} \otimes \sigma^{\beta}_{j}
\otimes ...]$, with $\alpha, \beta, ... = x, y, z,$ and $i, j, ... = 1,..., N$. By
Trotter expanding the associated time-evolution operator $U_t \approx (\Pi_{(i, j, ...)}
U_{(i, j, ...)})^n$, where $U_{(i, j, ...)} = \exp(-i \frac{t}{n} g_{(i, j, ...)}
[\sigma^{\alpha}_{i} \otimes \sigma^{\beta}_{j} \otimes ...])$ we find that fermionic
dynamics relies on the robust implementation of the gates $U_{(i, j, ...)}$. Note that
the latter can be achieved by a single mediator~\cite{Nielsen, Casanova12, Lamata14}
which adapts to our protocol.

Any unitary gate between different nuclei can
be implemented as $U_{q_i, \{ q_n \}}={\rm SWAP}_{q_e, q_i}
\ \tilde{U}_{q_e, \{ q_n\}} \ {\rm SWAP}_{q_e, q_i},$ where $q_e$ labels the electron
spin while $q_j$ the nuclear spins, $\tilde{U}_{q_e, \{ q_n\}}$
represents a gate between the electron spin and some set of $n$ qubits
$\{ q_n \}$ that does not contain the ${i}$-th qubit. When applied to some
initial state we find $U_{q_i, \{ q_n \}} |\psi_e\rangle |\psi_i\rangle |\phi_{\{q_n\}}\rangle
=|\psi_e\rangle \tilde{U}_{q_i, \{ q_n\}} |\psi_i\rangle |\phi_{\{q_n\}}\rangle$. This
procedure can be repeated  to find 
$|\psi_e\rangle \tilde{U}_{q_{i_j}, \{ q_{n_j}\}} ... \tilde{U}_{q_{i_2}, \{ q_{n_2}\}}
\tilde{U}_{q_{i_1}, \{ q_{n_1}\}} |\phi_{\{q_N\}}\rangle$, that corresponds to a set of
operations applied on the nuclei where $N=n+1$, and $n_j$, $q_{i_j}$
label different nuclear spin sets and spin targets respectively.

Quantum algorithms as outlined above assume a pure initial state of the entire
quantum computer. Remarkably, however, there are models of quantum computation that 
achieve  computational advantages with minimal coherence. Indeed, a 
single pure qubit (in our case the easily polarisable NV electron spin) 
executing a controlled unitary operation $|0\rangle\langle 0|\otimes \id + |1\rangle\langle 1|\otimes U$ 
on a maximally mixed register (the $^{13}$C nuclei)
can be used to obtain the trace of the unitary $U$ efficiently after a single measurement 
of the control qubit. This setting realises the DQC1 protocol \cite{KnillL1998} which can be shown to achieve computational advantages over 
classical computation and allow for the execution of complex algorithms such as 
Shor's factoring algorithm \cite{ParkerP2000, ParkerP2002}. Furthermore, the DQC1 model
can be adapted to obtain Heisenberg scaling in metrology tasks which ties in well with envisaged spin sensing applications \cite{BoixoS08}. 

{\em Conclusions --}  We  provide a robust scheme to perform
highly-selective single- and two-qubit gates into different nuclear
spins while efficiently eliminate the effect of internuclear interactions and the noise on the mediator. Our method does not
requires strongly coupled spins representing  an scalable procedure for general
quantum computing purposes.

{\em Acknowledgements --} This work was supported by the Alexander von Humboldt Foundation, the
ERC Synergy grant BioQ, the EU projects DIADEMS, SIQS, EQUAM and HYPERDIAMOND as well as the
DFG via the SFB TRR/21 and the SPP 1601.

\pagebreak
\widetext
\begin{center}
\textbf{ \large Supplemental Material: \\Noise-resilient Quantum Computing with a Nitrogen-Vacancy Center and Nuclear Spins}
\end{center}
\setcounter{equation}{0}
\setcounter{figure}{0}
\setcounter{table}{0}
\makeatletter
\renewcommand{\theequation}{S\arabic{equation}}
\renewcommand{\thefigure}{S\arabic{figure}}
\renewcommand{\bibnumfmt}[1]{[S#1]}
\renewcommand{\citenumfont}[1]{S#1}

\section{Hamiltonian model under decoupling radio-frequency field }
Here we show how to achieve the effective Hamiltonian in Eq.~$(2)$ on the main text. In a rotating frame with respect to the electronic free energy terms $D S_z^2 - \gamma_e B_zS_z$ we have ($\hbar = 1$)
\begin{equation}\label{sup:root}
H = -\sum_j \omega \ I_j^z + \frac{m_s}{2} [F(t) \ \sigma_z + \mathbb{I}]\sum_j \vec{A}_j \cdot \vec{I}_j +2 \Omega \cos{(\omega_{\rm rf} t)}  \sum_j   \vec{I}_j \cdot \hat{n} + \tilde{H}_{\rm c} + H_{nn}.
\end{equation}
Here $\omega = \gamma_n B_z$ with $\gamma_n = 2 \pi \times 10.705  \ \frac{\rm MHz}{\rm T}$ being the $^{13}$C gyromagnetic ratio, $m_s = \pm 1$ see~\cite{Casanova15}, $\vec{A}_{j}$ is the hyperfine vector for the $j$-th nucleus, $2\Omega = \gamma_n B_{\rm d}$, $\sigma_z = |m_s\rangle \langle m_s| -  | 0 \rangle \langle 0|$ denotes the electron spin Pauli operator, $\tilde{H}_{\rm c}$ gives rise to the action of the rf control field, $F(t)=1$, or $F(t)=-1$ (depending on whether an  even, $F(t)=1$, or odd, $F(t)=-1$, number of  decoupling pulses have been applied) is the modulation function and $H_{\rm nn}$  is
\begin{equation}
    H_{\text{nn}}=\sum_{j>k}\frac{\mu_{0}}{2}\frac{\gamma_n^2}{r_{j,k}^{3}} \left[\vec{I}_{j}\cdot\vec{I}_{k}-\frac{3(\vec{I}_{j}\cdot\vec{r}_{j,k})
    (\vec{r}_{j,k}\cdot\vec{I}_{k})}{r_{j,k}^{2}}\right],
\end{equation}
where $|\vec{r}_{j,k}|$ is the distance between the $j$-, and $k$-th nuclei. Additionally and without loss of generality we consider  $\hat{n} = (n^{z^{\perp}} , 0 , n^z)$ with $z^{\perp}\equiv \hat{x}$ being a direction orthogonal to $\hat{z}$.

Our analysis will be performed in a counter rotating frame that allows to a more detailed description of the resonances structure in Eq.~(\ref{sup:root}) than the one that can be obtained by merely applying the rotating wave approximation. More specifically, in order to take into account possible energy deviations coming from Bloch-Siegert shifts~\cite{Allen}  it is convenient to move into an interaction picture w.r.t.  $\omega_{\rm rf} \sum_j I_j^z$. This yields 
\begin{equation}
\begin{array}{ccc}
H  &=& \sum_j \bigg[  \Omega_x I_j^x  - (\omega + \omega_{\rm rf}) I_j^z \  \bigg] + \Omega_x \sum_j \bigg[I_j^+ e^{i 2\omega_{\rm rf} t}  + I_j^- e^{-i 2\omega_{\rm rf} t}\bigg] + 2 \Omega_z \cos{(\omega_{\rm rf} t)}\sum_j  I_j^z \vspace{0.25 cm} \\ \nonumber 
&&+ \frac{m_s}{2} [F(t) \ \sigma_z + \mathbb{I}]\sum_j  e^{i \omega_{\rm rf} I_j^z t} \vec{A}_j \cdot \vec{I}_j e^{-i \omega_{\rm rf} I_j^z t} +  e^{it  \sum_j\omega_{\rm rf} I_j^z} \tilde{H}_{\rm c} e^{-it  \sum_j\omega_{\rm rf} I_j^z}  +\tilde{H}_{nn}.
\end{array}
\end{equation}
Here, $\Omega_{x, z} = \Omega \ n^{z^{\perp}, z}$ and  $\tilde{H}_{nn}$ reads
\begin{equation}
\tilde{H}_{nn} = e^{it \sum_j \omega_{\rm rf} I_j^z}  H_{nn} \ e^{-it  \sum_j \omega_{\rm rf} I_j^z} \approx \sum_{j>k}\frac{\mu_{0}}{2}\frac{\gamma_{n}^2}{r_{j,k}^{3}} \left[1 - 3 (n_{j,k}^z)^2 \right] \left[I^z_{j} \ I^z_{k} - \frac{1}{2} (I_j^{x} I_k^{y} + I_j^{y} I_k^{x})\right].
\end{equation}
By defining  $ \bigg[  \Omega_x I_j^x  - (\omega + \omega_{\rm rf}) I_j^z \  \bigg]  = \tilde{\omega} \ \tilde{n} \cdot\vec{I}_j$, where $\tilde{\omega}  = \sqrt{(\omega + \omega_{\rm rf})^2 + \Omega_x^2 }$ and $\tilde{n} = (\frac{\Omega_x}{\tilde\omega}, 0,  -\frac{\omega + \omega_{\rm rf}}{\tilde\omega})$, we can write 
\begin{equation}
\begin{array}{ccc}
H  &=&  \tilde{\omega}\sum_j  \tilde{n} \cdot \vec{I}_j  + \Omega_x \sum_j \bigg[I_j^+ e^{i 2\omega_{\rm rf} t}  + I_j^- e^{-i 2\omega_{\rm rf} t}\bigg] + 2 \Omega_z \cos{(\omega_{\rm rf} t)}\sum_j  I_j^z\vspace{0.2cm}\\ \nonumber &&+ \frac{m_s}{2} [F(t) \ \sigma_z + \mathbb{I}]\sum_j  e^{i \omega_{\rm rf} I_j^z t} \vec{A}_j \cdot \vec{I}_j e^{-i \omega_{\rm rf} I_j^z t} +e^{it  \sum_j\omega_{\rm rf} I_j^z} \tilde{H}_{\rm c} e^{-it  \sum_j\omega_{\rm rf} I_j^z} +  \tilde{H}_{nn}.
\end{array}
\end{equation}

In order to achieve internuclear decoupling the action of a magnetic field oriented into an specific angle is required. This condition is known as the magic angle condition~\cite{Lee65} and it can be addressed in our formalism by expressing $\tilde{\omega} = \Delta  + \xi$ and moving to a rotating frame w.r.t. $\xi \sum_j  \tilde{n} \cdot \vec{I}_j$. By using the identity 
\begin{equation}\label{sup:identity}
e^{i\vec{I}_{j}\cdot\hat{l}\phi}\vec{I}_{j}\cdot\vec{b}\ e^{-i\vec{I}_{j}\cdot\hat{l}\phi}=\vec{I}_{j}\cdot[(\vec{b}-\vec{b}\cdot\hat{l}\hat{l})\cos\phi-\hat{l}\times\vec{b}\sin\phi+\vec{b}\cdot\hat{l}\hat{l}],
\end{equation}  
and under the resonance condition $\xi = 2\omega_{\rm rf}$ we find 
\begin{eqnarray}\label{sup:mesh}
H =&& \sum_j \bigg\{ \big[ \Delta  \tilde{n}_x + \frac{\Omega_x}{2} (1 - \tilde{n}_x^2 - \tilde{n}_z)\big] \ I_j^x + \big[\Delta \tilde{n}_z + \frac{\Omega_x}{2}\tilde{n}_x (1 - \tilde{n}_z)\big] \  I_j^z \bigg\}\nonumber\\
&&+ \frac{m_s}{2} [F(t) \ \sigma_z + \mathbb{I}]\sum_j e^{i \xi  \tilde{n} \cdot \vec{I}_j t} e^{i \omega_{\rm rf} I_j^z t} \vec{A}_j \cdot \vec{I}_j e^{-i \omega_{\rm rf} I_j^z t} e^{-i \xi  \tilde{n} \cdot \vec{I}_j t}\nonumber\\
&&+ \ e^{i \xi  \sum_j\tilde{n} \cdot \vec{I}_j t} e^{it  \sum_j\omega_{\rm rf} I_j^z} \tilde{H}_{\rm c} e^{-it  \sum_j\omega_{\rm rf} I_j^z}  e^{-i \xi  \sum_j\tilde{n} \cdot \vec{I}_j t} + e^{i \xi  \sum_j\tilde{n} \cdot \vec{I}_j t}  \tilde{H}_{nn} e^{-i \xi  \sum_j\tilde{n} \cdot \vec{I}_j t}.
\end{eqnarray}
In the previous equation we have  eliminated the term $2 \Omega_z \cos{(\omega_{\rm rf} t)}\sum_j  I_j^z$ that it is not  compensated with any other oscillating variable, see section~\ref{secfidel} for further considerations about the fidelity of this effective Hamiltonian. Note that the resonance condition $2\omega_{\rm rf} = \xi$ implies  
\begin{equation}
\Delta =  \tilde{\omega} - 2\omega_{\rm rf}  = \sqrt{(\omega + \omega_{\rm rf})^2 + \Omega^2_x} - 2\omega_{\rm rf}=  (\omega + \omega_{\rm rf}) \sqrt{1+ \bigg(\frac{\Omega_x}{\omega + \omega_{\rm rf}} \bigg)^2} - 2\omega_{\rm rf} \approx  \omega - \omega_{\rm rf} + \frac{1}{2} \frac{\Omega_x^2}{\omega + \omega_{\rm rf}},
\end{equation} 
where it can be explicitly seen an energy shift proportional to $\frac{\Omega_x^2}{\omega + \omega_{\rm rf}}$, see \cite{Allen}.

The second line of Eq.~(\ref{sup:mesh}) can be simplified to
\begin{equation}
 \frac{m_s}{2} F(t) \ \sigma_z\sum_j e^{i \xi  \tilde{n} \cdot \vec{I}_j t} e^{i \omega_{\rm rf} I_j^z t} \vec{A}_j \cdot \vec{I}_j e^{-i \omega_{\rm rf} I_j^z t} e^{-i \xi  \tilde{n} \cdot \vec{I}_j t}  +  \frac{m_s}{2}   \sum_j A_j^z  (\hat{z} \cdot \tilde{n}) \  \vec{I}_j  \cdot  \tilde{n},
\end{equation}
where we have eliminated fast rotating terms. This gives rise to 

\begin{eqnarray}\label{sup:demons}
H = &&\sum_j \bigg\{ \big[ \Delta  \tilde{n}_x + \frac{\Omega_x}{2} (1 - \tilde{n}_x^2 - \tilde{n}_z) + \frac{m_s}{2} \tilde{n}_z \ \tilde{n}_x \ A_j^z \big] \ I_j^x + \big[\Delta \tilde{n}_z + \frac{\Omega_x}{2}\tilde{n}_x (1 - \tilde{n}_z)+ \frac{m_s}{2}  \ \tilde{n}^2_z \ A_j^z \big]  \ I_j^z  \bigg\}\nonumber\\
&&+ \frac{m_s}{2} F(t) \ \sigma_z \sum_j e^{i \xi  \tilde{n} \cdot \vec{I}_j t} e^{i \omega_{\rm rf} I_j^z t} \vec{A}_j \cdot \vec{I}_j e^{-i \omega_{\rm rf} I_j^z t} e^{-i \xi  \tilde{n} \cdot \vec{I}_j t}\nonumber\\
&&+ \ e^{i \xi  \sum_j\tilde{n} \cdot \vec{I}_j t} e^{it  \sum_j\omega_{\rm rf} I_j^z} \tilde{H}_{\rm c} e^{-it  \sum_j\omega_{\rm rf} I_j^z}  e^{-i \xi  \sum_j\tilde{n} \cdot \vec{I}_j t}+ e^{i \xi  \sum_j\tilde{n} \cdot \vec{I}_j t}  \tilde{H}_{nn} e^{-i \xi  \sum_j\tilde{n} \cdot \vec{I}_j t},
\end{eqnarray}
where $\tilde{n}_x = \tilde{n} \cdot \hat{x}$, and  $\tilde{n}_z = \tilde{n} \cdot \hat{z}$.

The elimination of the internuclear interactions  happens if we match the magic-angle condition~\cite{Lee65} which in our case corresponds to 
\begin{equation}
  \big[ \Delta \tilde{n}_x + \frac{\Omega_x}{2} (1 - \tilde{n}_x^2 - \tilde{n}_z) + \frac{m_s}{2} \tilde{n}_z \ \tilde{n}_x \ A_j^z \big] = \sqrt{2} \ \big[\Delta \tilde{n}_z + \frac{\Omega_x}{2}\tilde{n}_x (1 - \tilde{n}_z) + \frac{m_s}{2} \ \tilde{n}^2_z \ A_j^z\big].
\end{equation}
Because $A_j^z$ takes different values for each nuclei the above equality does not hold for any $j$, however it can be approximately fulfilled if $\Delta$ takes a  value  such that 
\begin{equation}\label{sup:manglerestriction}
 | A_j^z| \ll |2 \Delta|, \ \ \forall j,
\end{equation} 
This issue simplifies the magic-angle condition that now reads
\begin{equation}\label{sup:mangle}
   \tilde{n}_x + \frac{\Omega_x}{2 \Delta} (1 - \tilde{n}_x^2 - \tilde{n}_z)  = \sqrt{2} \ \big[ \tilde{n}_z + \frac{\Omega_x}{2 \Delta}\tilde{n}_x (1 - \tilde{n}_z) ].
\end{equation}
Note that the above equation establishes a relation between the externally controllable parameters $\Omega$, $\hat{n}$, and $\Delta$.

In this manner the Hamiltonian is 
\begin{eqnarray}\label{sup:tunable}
H = \sum_j \omega_j \ \hat{n}_j \cdot \vec{I}_j  +  \frac{m_s}{2}F(t) \ \sigma_z \sum_j e^{i \xi  \tilde{n} \cdot \vec{I}_j t} e^{i \omega_{\rm rf} I_j^z t} \vec{A}_j \cdot \vec{I}_j e^{-i \omega_{\rm rf} I_j^z t} e^{-i \xi  \tilde{n} \cdot \vec{I}_j t}\nonumber\\
 + e^{i \xi  \sum_j\tilde{n} \cdot \vec{I}_j t} e^{it  \sum_j\omega_{\rm rf} I_j^z} \tilde{H}_{\rm c} e^{-it  \sum_j\omega_{\rm rf} I_j^z}  e^{-i \xi  \sum_j\tilde{n} \cdot \vec{I}_j t}+ e^{i \xi  \sum_j\tilde{n} \cdot \vec{I}_j t}  \tilde{H}_{nn} e^{-i \xi  \sum_j\tilde{n} \cdot \vec{I}_j t}.
\end{eqnarray}
The above expression contains a resonance frequency, $\omega_j$,  depending on the nuclear register. We can exploit this issue  to generate single-qubit addressing and implement entangling operations between the electron spin and some specific nucleus  as well as  single qubit rotations. More specifically  that frequency is  

\begin{equation}
\omega_j = |\Delta| \sqrt{ \bigg[\delta + \frac{m_s}{2 \Delta}  \tilde{n}_z \ \tilde{n}_x \ A_j^z \bigg]^2 + \bigg[\frac{\delta}{\sqrt{2}} + \frac{m_s}{2 \Delta}   \ \tilde{n}^2_z \ A_j^z \bigg]^2}
\end{equation}
with $\delta =  \tilde{n}_x + \frac{\Omega_x}{2 \Delta} (1 - \tilde{n}_x^2 - \tilde{n}_z)$. The $j$-th nuclear rotation axis reads
\begin{equation}
\hat{n}_j = \bigg( \frac{\Delta\delta + \frac{m_s}{2}  \tilde{n}_z \ \tilde{n}_x \ A_j^z}{\omega_j}, 0, \frac{\frac{\Delta\delta}{\sqrt{2}} + \frac{m_s}{2}   \ \tilde{n}^2_z \ A_j^z}{\omega_j} \bigg) 
\end{equation}
An additional change of interaction picture w.r.t. $ \sum_j \omega_j \ \hat{n}_j \cdot \vec{I}_j$ provides with 

\begin{eqnarray}
H &= & \frac{m_s}{2}F(t) \ \sigma_z \sum_j e^{i  \omega_j \ \hat{n}_j \cdot \vec{I}_j t} \ e^{i \xi  \tilde{n} \cdot \vec{I}_j t} \ e^{i \omega_{\rm rf} I_j^z t} \ \vec{A}_j \cdot \vec{I}_j \ e^{-i \omega_{\rm rf} I_j^z t} \ e^{-i \xi  \tilde{n} \cdot \vec{I}_j t} \  e^{-i  \omega_j \ \hat{n}_j \cdot \vec{I}_j t} \nonumber\\
 &+& e^{i \sum_j \omega_j \ \hat{n}_j \cdot \vec{I}_j t} \ e^{i \xi  \sum_j\tilde{n} \cdot \vec{I}_j t} \ e^{it  \sum_j\omega_{\rm rf} I_j^z} \tilde{H}_{\rm c} \ e^{-it  \sum_j\omega_{\rm rf} I_j^z} \ e^{-i \xi  \sum_j\tilde{n} \cdot \vec{I}_j t} \ e^{-i \sum_j \omega_j \ \hat{n}_j \cdot \vec{I}_j t} \nonumber\\
 &+& e^{i \sum_j \omega_j \ \hat{n}_j \cdot \vec{I}_j t} \ e^{i \xi  \sum_j\tilde{n} \cdot \vec{I}_j t} \  \tilde{H}_{nn} \ e^{-i \xi  \sum_j\tilde{n} \cdot \vec{I}_j t} \ e^{-i \sum_j \omega_j \ \hat{n}_j \cdot \vec{I}_j t}.
\end{eqnarray}
The last line of the previous equation, i.e. the one including the internuclear interactions, can be treated as follows. First, for large values of $B_z$  we have $\tilde{n} \approx \hat{z}$ and consequently 
\begin{equation}
\ e^{i \xi  \sum_j\tilde{n} \cdot \vec{I}_j t} \  \tilde{H}_{nn} \ e^{-i \xi  \sum_j\tilde{n} \cdot \vec{I}_j t} \approx \tilde{H}_{nn}.
\end{equation}
Then we can average out the remaining expression, $e^{i \sum_j \omega_j \ \hat{n}_j \cdot \vec{I}_j t} \   \tilde{H}_{nn} \ e^{-i \sum_j \omega_j \ \hat{n}_j \cdot \vec{I}_j t}$, because of the magic angle condition in Eq.~(\ref{sup:mangle})  and  the easily achievable requirement 
\begin{equation}
\big|\frac{\mu_{0}\gamma^2_{n}}{2|\vec{r}_{j,k}|^{3}}\big|   \ll |\omega_j|  
\end{equation}
Note that for the diamond lattice the lowest Carbon-Carbon distance is 
$r_0 \approx 1.54 \  \mathring{\rm A}$, therefore nuclei located at that distance have the highest value of the coupling  coefficient $\frac{\mu_{0}\gamma^2_{j}}{2|\vec{r}_{j,k}|^{3}}$. Consequently  we need a set of frequencies $\omega_j$ such that 
\begin{equation}
2 \ \mbox{kHz}  \ll |\omega_j| \end{equation}
In this manner  internuclear interactions can be approximately neglected, note that in addition to the previous condition we are also restricted by the applicability of Eq.~(\ref{sup:manglerestriction}). 

Hence we can simplify the Hamiltonian to the following expression 

\begin{eqnarray}\label{sup:prior}
H &= & \frac{m_s}{2}F(t) \ \sigma_z \sum_j e^{i  \omega_j \ \hat{n}_j \cdot \vec{I}_j t} \ e^{i \xi  \tilde{n} \cdot \vec{I}_j t} \ e^{i \omega_{\rm rf} I_j^z t} \ \vec{A}_j \cdot \vec{I}_j \ e^{-i \omega_{\rm rf} I_j^z t} \ e^{-i \xi  \tilde{n} \cdot \vec{I}_j t} \  e^{-i  \omega_j \ \hat{n}_j \cdot \vec{I}_j t} \nonumber\\
 &+& e^{i \sum_j \omega_j \ \hat{n}_j \cdot \vec{I}_j t} \ e^{i \xi  \sum_j\tilde{n} \cdot \vec{I}_j t} \ e^{it  \sum_j\omega_{\rm rf} I_j^z} \tilde{H}_{\rm c} \ e^{-it  \sum_j\omega_{\rm rf} I_j^z} \ e^{-i \xi  \sum_j\tilde{n} \cdot \vec{I}_j t} \ e^{-i \sum_j \omega_j \ \hat{n}_j \cdot \vec{I}_j t}. 
 \end{eqnarray}

With the help of the identity in Eq.~(\ref{sup:identity}) and writing the control Hamiltonian $H_{\rm c}$ as 
\begin{equation}
H_{\rm c} = 2\lambda\cos(\omega_{c} + \phi_c) \sum_j \vec{I}_j \cdot \hat{n}_c, 
\end{equation}
Eq.~(\ref{sup:prior})  can be fully developed providing a complete map to the resonances structure of the system under the action of the decoupling field. This is
\begin{eqnarray}\label{sup:super}
H  = \frac{m_s}{2}F(t) \ \sigma_z \sum_j &\bigg\{& \vec{I}_j\cdot\bigg[(\vec{\alpha}_{1,j} - \vec{\alpha}_{1,j}\cdot\hat{n}_j \ \hat{n}_j) \cos{(\omega_j t)} - (\hat{n}_j \times \vec{\alpha}_{1,j}) \sin{(\omega_j t)}  + \vec{\alpha}_{1,j}\cdot\hat{n}_j \ \hat{n}_j \bigg] \cos{(\xi t)} \cos{(\omega_{\rm rf } t)}  \nonumber\\
&-&  \vec{I}_j\cdot\bigg[(\vec{\alpha}_{2,j} - \vec{\alpha}_{2,j}\cdot\hat{n}_j \ \hat{n}_j) \cos{(\omega_j t)} - (\hat{n}_j \times \vec{\alpha}_{2,j}) \sin{(\omega_j t)}  + \vec{\alpha}_{2,j}\cdot\hat{n}_j \ \hat{n}_j \bigg] \sin{(\xi t)} \cos{(\omega_{\rm rf } t)} \nonumber\\
&+&  \vec{I}_j\cdot\bigg[(\vec{\alpha}_{3,j} - \vec{\alpha}_{3,j}\cdot\hat{n}_j \ \hat{n}_j) \cos{(\omega_j t)} - (\hat{n}_j \times \vec{\alpha}_{3,j}) \sin{(\omega_j t)}  + \vec{\alpha}_{3,j}\cdot\hat{n}_j \ \hat{n}_j \bigg] \cos{(\omega_{\rm rf}  t)} \nonumber\\
&-&\vec{I}_j\cdot\bigg[(\vec{\beta}_{1,j} - \vec{\beta}_{1,j}\cdot\hat{n}_j \ \hat{n}_j) \cos{(\omega_j t)} - (\hat{n}_j \times \vec{\beta}_{1,j}) \sin{(\omega_j t)}  + \vec{\beta}_{1,j}\cdot\hat{n}_j \ \hat{n}_j \bigg] \cos{(\xi t)} \sin{(\omega_{\rm rf } t)}\nonumber\\
&+&\vec{I}_j\cdot\bigg[(\vec{\beta}_{2,j} - \vec{\beta}_{2,j}\cdot\hat{n}_j \ \hat{n}_j) \cos{(\omega_j t)} - (\hat{n}_j \times \vec{\beta}_{2,j}) \sin{(\omega_j t)}  + \vec{\beta}_{2,j}\cdot\hat{n}_j \ \hat{n}_j \bigg] \sin{(\xi t)} \sin{(\omega_{\rm rf } t)}\nonumber\\
&-&\vec{I}_j\cdot\bigg[(\vec{\beta}_{3,j} - \vec{\beta}_{3,j}\cdot\hat{n}_j \ \hat{n}_j) \cos{(\omega_j t)} - (\hat{n}_j \times \vec{\beta}_{3,j}) \sin{(\omega_j t)}  + \vec{\beta}_{3,j}\cdot\hat{n}_j \ \hat{n}_j \bigg] \sin{(\omega_{\rm rf} t)} \nonumber\\
&+&\vec{I}_j\cdot\bigg[(\vec{\gamma}_{1,j} - \vec{\gamma}_{1,j}\cdot\hat{n}_j \ \hat{n}_j) \cos{(\omega_j t)} - (\hat{n}_j \times \vec{\gamma}_{1,j}) \sin{(\omega_j t)}  + \vec{\gamma}_{1,j}\cdot\hat{n}_j \ \hat{n}_j \bigg] \cos{(\xi t)} \nonumber\\
&-&\vec{I}_j\cdot\bigg[(\vec{\gamma}_{2,j} - \vec{\gamma}_{2,j}\cdot\hat{n}_j \ \hat{n}_j) \cos{(\omega_j t)} - (\hat{n}_j \times \vec{\gamma}_{2,j}) \sin{(\omega_j t)}  + \vec{\gamma}_{2,j}\cdot\hat{n}_j \ \hat{n}_j \bigg] \sin{(\xi t)} \nonumber\\
&+&\vec{I}_j\cdot\bigg[(\vec{\gamma}_{3,j} - \vec{\gamma}_{3,j}\cdot  \hat{n}_j \ \hat{n}_j) \cos{(\omega_j t)} - (\hat{n}_j \times \vec{\gamma}_{3,j}) \sin{(\omega_j t)}  + \vec{\gamma}_{3,j}\cdot\hat{n}_j \ \hat{n}_j \bigg] \  \  \ \bigg\}\nonumber\\
+2\lambda\cos(\omega_{c} + \phi_c) \sum_j &\bigg\{&  \vec{I}_j\cdot\bigg[(\vec{m}_{1,1} - \vec{m}_{1,1}\cdot\hat{n}_j \ \hat{n}_j) \cos{(\omega_j t)} - (\hat{n}_j \times \vec{m}_{1,1}) \sin{(\omega_j t)}  + \vec{m}_{1,1}\cdot\hat{n}_j \ \hat{n}_j \bigg] \cos{(\xi t)} \cos{(\omega_{\rm rf } t)} \nonumber\\
&-&  \vec{I}_j\cdot\bigg[(\vec{m}_{1,2} - \vec{m}_{1,2}\cdot\hat{n}_j \ \hat{n}_j) \cos{(\omega_j t)} - (\hat{n}_j \times \vec{m}_{1,2}) \sin{(\omega_j t)}  + \vec{m}_{1,2}\cdot\hat{n}_j \ \hat{n}_j \bigg] \sin{(\xi t)} \cos{(\omega_{\rm rf } t)} \nonumber\\
&+&  \vec{I}_j\cdot\bigg[(\vec{m}_{1,3} - \vec{m}_{1,3}\cdot\hat{n}_j \ \hat{n}_j) \cos{(\omega_j t)} - (\hat{n}_j \times \vec{m}_{1,3}) \sin{(\omega_j t)}  + \vec{m}_{1,3}\cdot\hat{n}_j \ \hat{n}_j \bigg] \cos{(\omega_{\rm rf}  t)} \nonumber\\
&-&  \vec{I}_j\cdot\bigg[(\vec{m}_{2,1} - \vec{m}_{2,1}\cdot\hat{n}_j \ \hat{n}_j) \cos{(\omega_j t)} - (\hat{n}_j \times \vec{m}_{2,1}) \sin{(\omega_j t)}  + \vec{m}_{2,1}\cdot\hat{n}_j \ \hat{n}_j \bigg] \cos{(\xi t)}\sin{(\omega_{\rm rf}  t)} \nonumber\\
&+&  \vec{I}_j\cdot\bigg[(\vec{m}_{2,2} - \vec{m}_{2,2}\cdot\hat{n}_j \ \hat{n}_j) \cos{(\omega_j t)} - (\hat{n}_j \times \vec{m}_{2,2}) \sin{(\omega_j t)}  + \vec{m}_{2,2}\cdot\hat{n}_j \ \hat{n}_j \bigg] \cos{(\xi t)}\sin{(\omega_{\rm rf}  t)} \nonumber\\
&-&  \vec{I}_j\cdot\bigg[(\vec{m}_{2,3} - \vec{m}_{2,3}\cdot\hat{n}_j \ \hat{n}_j) \cos{(\omega_j t)} - (\hat{n}_j \times \vec{m}_{2,3}) \sin{(\omega_j t)}  + \vec{m}_{2,3}\cdot\hat{n}_j \ \hat{n}_j \bigg] \sin{(\omega_{\rm rf}  t)} \nonumber\\
&+&  \vec{I}_j\cdot\bigg[(\vec{m}_{3,1} - \vec{m}_{3,1}\cdot\hat{n}_j \ \hat{n}_j) \cos{(\omega_j t)} - (\hat{n}_j \times \vec{m}_{3,1}) \sin{(\omega_j t)}  + \vec{m}_{3,1}\cdot\hat{n}_j \ \hat{n}_j \bigg] \cos{(\xi t)} \nonumber\\
&-&  \vec{I}_j\cdot\bigg[(\vec{m}_{3,2} - \vec{m}_{3,2}\cdot\hat{n}_j \ \hat{n}_j) \cos{(\omega_j t)} - (\hat{n}_j \times \vec{m}_{3,2}) \sin{(\omega_j t)}  + \vec{m}_{3,2}\cdot\hat{n}_j \ \hat{n}_j \bigg] \sin{(\xi t)} \nonumber\\
&+&  \vec{I}_j\cdot\bigg[(\vec{m}_{3,3} - \vec{m}_{3,3}\cdot\hat{n}_j \ \hat{n}_j) \cos{(\omega_j t)} - (\hat{n}_j \times \vec{m}_{3,3}) \sin{(\omega_j t)}  + \vec{m}_{3,3}\cdot\hat{n}_j \ \hat{n}_j \bigg] \ \ \bigg\}  \nonumber\\
\end{eqnarray}

where 
\begin{eqnarray}
\vec{\alpha}_{1,j} &=&  (\vec{A}_j - \vec{A}_j \cdot \hat{z} \ \hat{z}) - (\vec{A}_j - \vec{A}_j \cdot \hat{z} \ \hat{z}) \cdot \tilde{n} \ \tilde{n},\nonumber\\
\vec{\alpha}_{2,j} &=& \tilde{n} \ \times (\vec{A}_j - \vec{A}_j \cdot \hat{z} \ \hat{z}), \nonumber\\
\vec{\alpha}_{3,j} &=& (\vec{A}_j - \vec{A}_j \cdot \hat{z} \ \hat{z}) \cdot \tilde{n} \ \tilde{n},\nonumber\\
\vec{\beta}_{1,j} &=&  (\hat{z} \times \vec{A}_j) - (\hat{z} \times \vec{A}_j) \cdot \tilde{n} \ \tilde{n},\nonumber\\
\vec{\beta}_{2,j} &=& \tilde{n} \ \times (\hat{z} \times \vec{A}_j), \nonumber\\
\vec{\beta}_{3,j} &=& (\hat{z} \times \vec{A}_j) \cdot \tilde{n} \ \tilde{n},\nonumber\\
\vec{\gamma}_{1,j} &=&  (\vec{A}_j \cdot \hat{z} \ \hat{z}) - (\vec{A}_j \cdot \hat{z} \ \hat{z}) \cdot \tilde{n} \ \tilde{n},\nonumber\\
\vec{\gamma}_{2,j} &=& \tilde{n} \ \times (\vec{A}_j \cdot \hat{z} \ \hat{z}), \nonumber\\
\vec{\gamma}_{3,j} &=& (\vec{A}_j \cdot \hat{z}) \ (\hat{z} \cdot \tilde{n}) \ \tilde{n},
\end{eqnarray}
and 
\begin{eqnarray}
\vec{m}_{j,1} &=& \vec{n}_{c, j} - \vec{n}_{c, j}\cdot\tilde{n} \ \tilde{n},\nonumber\\
\vec{m}_{j,2} &=& \tilde{n} \times \vec{n}_{c, j}, \nonumber\\
\vec{m}_{j,3} &=&  \vec{n}_{c, j}\cdot\tilde{n} \ \tilde{n},\nonumber\\
\end{eqnarray}
with 
\begin{eqnarray}
\vec{n}_{c, 1} &=& \hat{n}_c -  \hat{n}_c \cdot \hat{z} \ \hat{z},\nonumber\\
\vec{n}_{c, 2} &=& \hat{z} \times \hat{n}_c, \nonumber\\
\vec{n}_{c, 3} &=& \hat{n}_c \cdot \hat{z} \ \hat{z}.
\end{eqnarray}

An inspection of Eq.~(\ref{sup:super}) together with the  condition $\xi = 2\omega_{\rm rf}$ reveals the existence of different resonance branches at $\omega_j$, $\omega_{\rm rf} \pm \omega_j$, and $3\omega_{\rm rf} \pm \omega_j$. We will restrict our analysis  to the branch that  resonates at  $\omega_j$ frequencies, i.e. those going with the $\vec{\gamma}_{3,j}$, and $\vec{m}_{3, 3}$ in Eq.~(\ref{sup:super}).  However the whole formalism can be applied to any other branch.

One can easily demonstrate that the set $\hat{x}_j$,  $\hat{y}_j$, and $\hat{z}_j$, where 
 $\vec{x}_j =  (\vec{\gamma}_{3,j} - \vec{\gamma}_{3,j}\cdot  \hat{n}_j \ \hat{n}_j)$, $\vec{y}_j=(\hat{n}_j \times \vec{\gamma}_{3,j})$, $\vec{z}_j=\hat{n}_j $ and $|\vec{x}_j| = |\vec{y}_j| =\big| (\vec{A}_j \cdot \hat{z})  (\hat{z} \cdot \tilde{n}) \sin{(\theta_{\tilde{n},  \hat{n}_j})} \big| = g_j$, $\theta_{\tilde{n},  \hat{n}_j}$ being the angle between the vectors $\tilde{n},  \hat{n}_j$,  constitutes a set of orthogonal and unitary vectors that allows to define the computational basis of our problem as follows
\begin{eqnarray}
I_j^x &=& \vec{I}_j \cdot \hat{x}_j,\nonumber\\
I_j^y &=& \vec{I}_j \cdot \hat{y}_j,\nonumber\\
I_j^z &=& \vec{I}_j \cdot \hat{z}_j.\nonumber\\
\end{eqnarray}
Additionally  we can express the set  $(\vec{m}_{3,3} - \vec{m}_{3,3}\cdot\hat{n}_j \ \hat{n}_j)$, $(\hat{n}_j \times \vec{m}_{3,3})$, and $\hat{n}_j$ in terms of the  previously defined vector basis. 

Despite the complexity of the Hamiltonian~(\ref{sup:super}) we can excite only one resonance branch by means of the robust AXY-8 sequence~\cite{Casanova15}  where the function $F(t)$ can be modulated reading $F(t) = \sum_{k>0} f_k \cos{(k \omega t)}$ with  $f_k$ and  $\omega$ being fully tunable parameters (see  section~\ref{symantisym} for a discussion of symmetric and antisymmetric  pulse arrangements of the AXY sequence). Then, if we set $F(t)$ with a period $\tau$ and  the control RF field  with a frequency  $\omega_c$  such that  $\left(\frac{k}{\tau}\right)$ and $\omega_c$   are both on the range of the resonant $\omega_j$ frequencies,  the Hamiltonian in Eq.~(\ref{sup:super}) is simplified to 
\begin{equation}\label{sup:ya}
H  = \frac{m_s}{2} F(t) \ \sigma_z  \sum_j g_j \bigg[I_j^x \cos{(\omega_j t)} -  I_j^y  \sin{(\omega_j t)}  \bigg] + 2\lambda\cos{(\omega_c t +\phi_c)}  \sum_j \vec{I}_j \cdot \bigg[ |\vec{\alpha}_j| \ \hat{\alpha}_j \cos{(\omega_j t)} -  |\vec{\beta}_j| \ \hat{\beta}_j \sin{(\omega_j t)}  \bigg],
\end{equation}
where $\vec{\alpha}_j$ and  $\vec{\beta}_j$ depends on the specific orientation of $\hat{n}_c$ with respect to the basis $\hat{x}_j$,  $\hat{y}_j$, and $\hat{z}_j$.

\section{Decoupling field  approximation}\label{secfidel}
In order to justify the decoupling field structure that gives rise to Eq.~(\ref{sup:tunable}) it is enough with considering the situation described by the expression
\begin{equation}\label{sup:real}
H =  - \omega \ I^z +   2 \Omega \  \vec{I} \cdot \hat{n} \cos{(\omega_{\rm rf} t)}. 
\end{equation}
\begin{figure}[t]
\includegraphics[width=0.6\columnwidth]{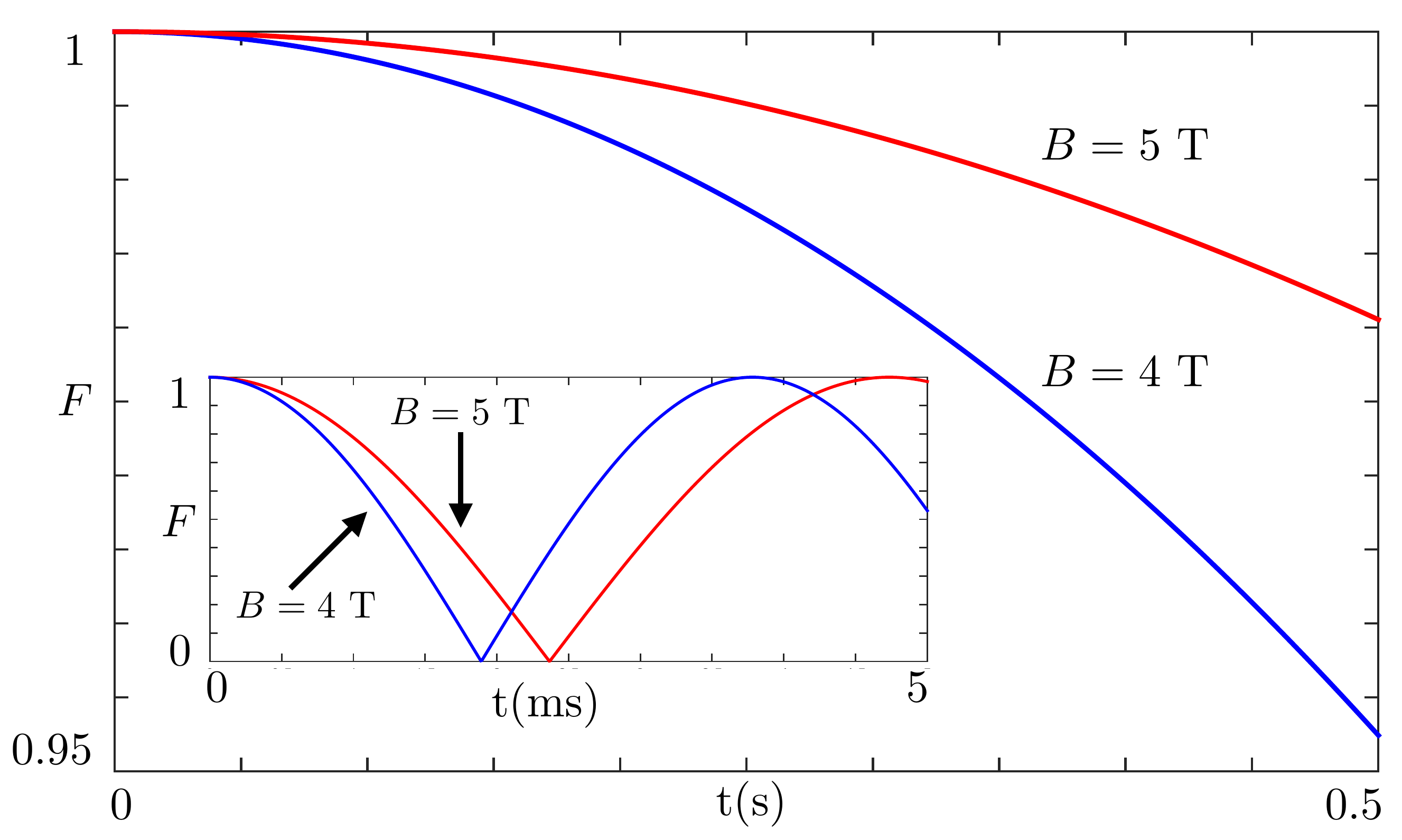}\caption{\label{sup:fidelity}(color online) fidelity values for the case $\Delta = 2\pi \times 200$ kHz and $\Omega $ calculated by using the magic-angle condition in Eq.~(\ref{sup:mangle}) for different values ($4$ and $5$ T) of the magnetic field in the $z$ direction. In both cases we assume an error in the radio frequency field alignment of $\pm 1.5^{\circ}$ in any direction considering $n_x =1$ as the ideal case. For these parameters the fidelity reaches values above $0.95$ for times up to $0.5$ seconds. In the inset we show the fidelity  when the RWA is applied. Note that in the latter the approximation given by the Hamiltonian in Eq.~(\ref{sup:RWA2}) is only valid for some hundreds of microseconds. }
\end{figure}
In the case of tuning $\omega_{\rm rf} = \omega$, the RWA gives rise to the following Hamiltonian in the interaction picture of $-\omega \ I^z$
\begin{equation}\label{sup:RWA}
H = \Omega_x I_x,
\end{equation}
or when $\omega_{\rm rf} = \omega - \Delta$ 
\begin{equation}\label{sup:RWA2}
H = -\Delta I_z +   \Omega_x I_x.
\end{equation}

However we can refine the RWA by proceeding as follows, in a rotating frame w.r.t. $\omega \ I^z$ (the counter rotating frame) we find 
\begin{equation}
H = \tilde{\omega} \ \tilde{n}  \cdot \vec{I}  + \Omega_x \  (I^{+} e^{i 2\omega t} + I^{-} e^{-i 2\omega t}) + 2\Omega_z \ I^z \cos{(\omega_{\rm rf} t)}.
\end{equation}
By writing $\tilde{\omega} = \Delta + \xi$ and neglecting the term $2\Omega_z \ I^z \cos{(\omega_{\rm rf} t)}$ that will be not compensated with other oscillating variable we find that, in the rotating frame of $\xi \ \tilde{n} \cdot \vec{I}$ and under the resonance condition $\xi = 2\omega_{\rm rf}$, the Hamiltonian reads
\begin{equation}\label{sup:effectivo}
H = \bigg[ \Delta \tilde{n}_x + \frac{\Omega_x}{2} (1 - \tilde{n}_x^2 - \tilde{n}_z)\bigg] I^x + \bigg[\Delta \tilde{n}_z + \frac{\Omega_x}{2}\tilde{n}_x (1 - \tilde{n}_z)\bigg]  I^z. 
\end{equation}

The above expression is only an approximation of the real Hamiltonian in Eq.~(\ref{sup:real}), however one can compute the fidelity between the propagators associated to  Eqs.~(\ref{sup:real}, \ref{sup:RWA})  and  Eqs.~(\ref{sup:real}, \ref{sup:effectivo})  as a function of time according to the definition 
\begin{equation}
F = \frac{|{\rm Tr}(A B^{\dag})|}{\sqrt{{\rm Tr}(A A^{\dag}) \ {\rm Tr}(B B^{\dag})}}
\end{equation}
where $A$ are $B$ are two general quantum operations. In Fig.~\ref{sup:fidelity} we show the fidelity of the time evolution operators generated by the Hamiltonians in Eqs.~(\ref{sup:real}, \ref{sup:effectivo})  for different values of the static magnetic field. The inset shows the fidelity when the RWA is invoked. An inspection of the plots shows an  improvement of almost three  orders of magnitude in the fidelity when working in the counter-rotating picture.

\section{Symmetric and anti-symmetric arrangement of AXY sequences}\label{symantisym}

\begin{figure}[t]
\includegraphics[width=0.7\columnwidth]{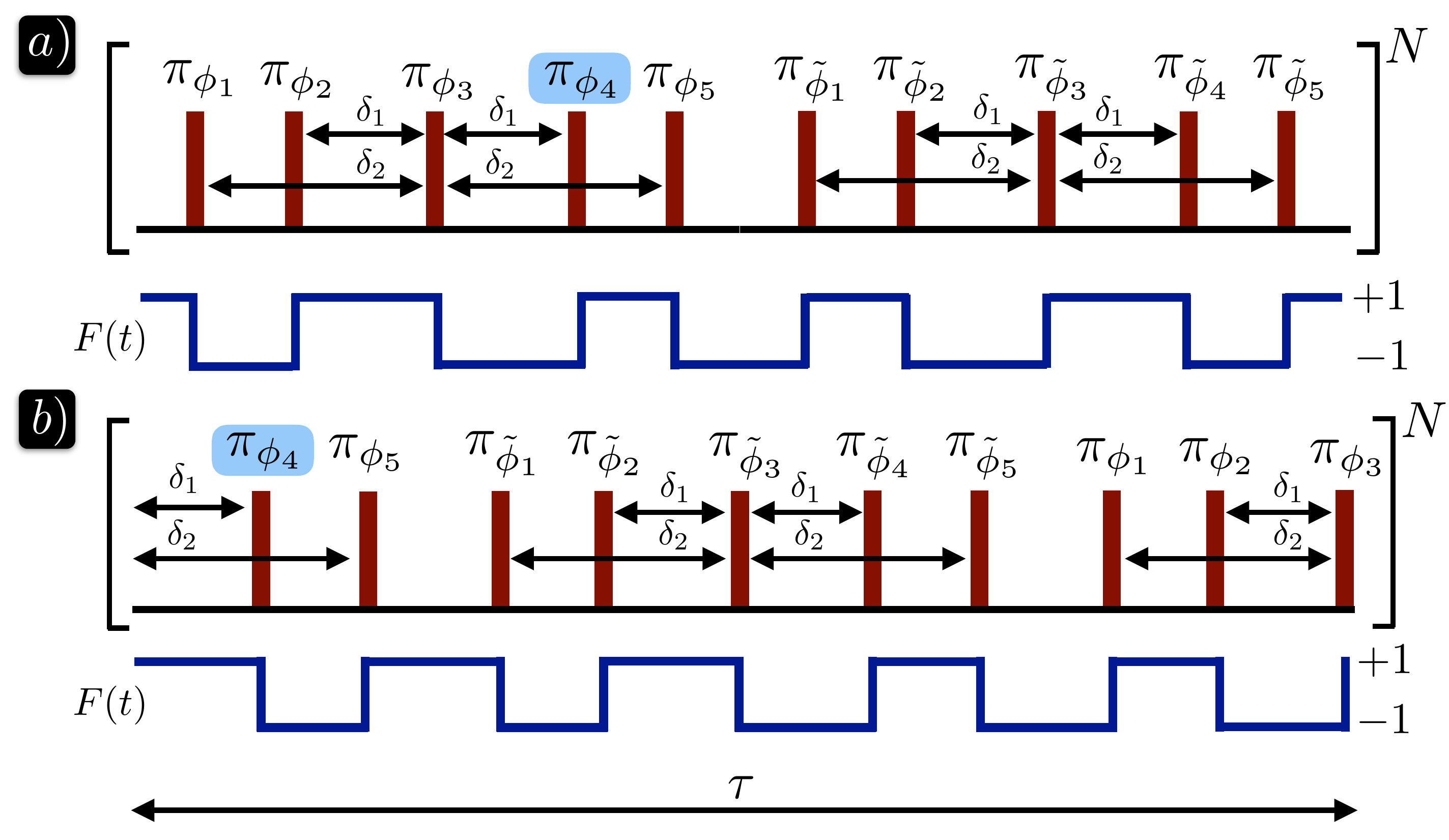}\caption{\label{sup:pulses}(color online) Different  pulse arrangements giving rise to symmetric a), and anti-symmetric b) AXY sequences. Behind each pulse sequence the associated modulation function, $F(t)$, is plotted  showing their even or odd character. The phases $\phi_j$, $\tilde{\phi}_j$  that determine the rotation axis in the X-Y plane are chosen such that  $\phi_j = [\frac{\pi}{6}, 0, \frac{\pi}{2}, 0,  \frac{\pi}{6}]$, and  $\tilde{\phi}_j = [\frac{\pi}{6}+ \frac{\pi}{2},  \frac{\pi}{2}, \frac{\pi}{2}+ \frac{\pi}{2},  \frac{\pi}{2},  \frac{\pi}{6}+ \frac{\pi}{2}]$.  Each pulse sequence is repeated $N$ times with the interpulse spacing  governed by the time-variables $\delta_1$, and $\delta_2$ that allow to arbitrarily tune the $f^{e}_k$, $f^{o}_k$ coefficients. The highlighted $\pi$-pulse in a), b) represents the link between the even an odd pulse sequences, i.e. the odd sequence is displayed considering the rotation around the axis $\phi_4$ as the first pulse.}
\end{figure}

In order to couple the electron spin with the $I_j^x$, or $I_j^y$ components of each individual nuclei, we can choose between the symmetric and anti-symmetric versions of the AXY sequence, see Fig.~\ref{sup:pulses}. While the sequence in Fig.~\ref{sup:pulses} a) produces an even modulation function 
$F(t)$ such that $F(t) = \sum_{k>0} f^{s}_k \cos{(k \omega t)}$, the arrangement of pulses in Fig.~\ref{sup:pulses} b) gives rise to   $F(t) = \sum_{k>0} f^{a}_k \sin{(k \omega t)}$ which presents an odd behavior. Note that the even an odd pulse sequences are simply related by a time-shift. More specifically, the $\pi$ pulse that rotates the electron spin along the axis defined by $\phi_4$ in Fig.~\ref{sup:pulses} a), i.e. in the even AXY sequence, is chosen as the first rotating pulse in the odd version of AXY. In both cases the coefficients $f^{s}_k$, $f^{a}_k$ can be arbitrarily tuned which provides with selective nuclear addressing and with high fidelity quantum gates (see discussion on the main text). In order to fit these pulse arrangements  with the standard XY-$8$ sequence we can consider, in a), the composite X pulse as the pulse-set displayed from $0$ to $\frac{\tau}{2}$ while the composite Y pulse is conformed by the pulses from $\frac{\tau}{2}$ to $\tau$. In this manner the symmetric AXY-$8$ sequence corresponds to the repeated application of the XYXYYXYX pulse arrangement. The anty-simmetric AXY-$8$ sequence can be built in the same way but taking into account that the composite X and Y pulses have to be constructed according to  Fig.~\ref{sup:pulses} b). 

\section{Decoupling Efficiency} 

The dynamics of interacting nuclear spins under the action of an static $B_z$ and a decoupling rf field is described by the Hamiltonian
\begin{eqnarray}
H &=& -\sum_j \omega I_j^z  + 2\Omega \cos{(\omega_{\rm rf} t)}  \sum_j  I^x_j  + \sum_{j>k}\frac{\mu_{0}}{2}\frac{\gamma_n^2}{r_{j,k}^{3}} \left[\vec{I}_{j}\cdot\vec{I}_{k}-\frac{3(\vec{I}_{j}\cdot\vec{r}_{j,k})(\vec{r}_{j,k}\cdot\vec{I}_{k})}{r_{j,k}^{2}}\right]\nonumber\\
    &\approx&-\sum_j \omega I_j^z  + 2\Omega \cos{(\omega_{\rm rf} t)}  \sum_j  I^x_j  + \sum_{j>k}\frac{\mu_{0}}{2}\frac{\gamma_n^2}{r_{j,k}^{3}} (1 - 3 \hat{r}_{j,k})
\left[I^z_j I^z_k -\frac{1}{2}(I^x_j I^x_k + I^y_j I^y_k )  \right].
\end{eqnarray}
Then, in a rotating frame w.r.t. $\omega_{\rm rf} \sum_j I_j^z$ such that $\omega_{\rm rf} = -\omega + \Delta$ we have 
\begin{equation}\label{sup:twonuclei}
H = -\Delta\sum_j  I_j^z  + \Omega  \sum_j  I^x_j  + \sum_{j>k}\lambda_{j,k}
\left[I^z_j I^z_k -\frac{1}{2}(I^x_j I^x_k + I^y_j I^y_k )  \right],
\end{equation}
where $\lambda_{j,k} = \frac{\mu_{0}}{2}\frac{\gamma_n^2}{r_{j,k}^{3}} (1 - 3 \hat{r}_{j,k})$. In order to simplify the explanation let us restrict to the simplest case of two interacting nuclei because the structure of the above Hamiltonian allows to straightforwardly extend the argument to many particles. In this manner we can consider the problem
\begin{equation}
H = \omega_1 \tilde{I}^z_1 +  \omega_2 \tilde{I}^z_2 +  \lambda_{1,2} \left[I^z_1 I^z_2 -\frac{1}{2}(I^x_1 I^x_2 + I^y_1 I^y_2 )  \right],
\end{equation}
where $\tilde{I}_{j}^z = \left[\frac{-\Delta}{\sqrt{\Delta^2 + \Omega^2 }}I_{j}^z + \frac{\Omega}{\sqrt{\Delta^2 + \Omega^2 }}I_{j}^x  \right]$, for $j=1,2$ and $\omega_1 = \omega_2 = \sqrt{\Delta^2 + \Omega^2 }$. Then, one can define a new base in terms of the $\tilde{I}^\alpha_j$ operators  as 
\begin{eqnarray}
\tilde{I}_j^z &=& e^{-i\theta I_j^y} \tilde{I}_j^z e^{i\theta I_j^y} =\cos{\theta} \ I_j^z + \sin{\theta} \ I_j^x,\nonumber\\
\tilde{I}_j^x &=& e^{-i\theta I_j^y} I_j^x e^{i\theta I_j^x} =\cos{\theta} \ I_j^x  - \sin{\theta} \ I_j^z,\nonumber\\
\tilde{I}_j^y &=&  I_j^y, 
\end{eqnarray}
or, conversely
\begin{eqnarray}
I_j^z &=& e^{i\theta \tilde{I}_j^y} \tilde{I}_j^z e^{-i\theta \tilde{I}_j^y} = \cos{\theta} \ \tilde{I}_j^z - \sin{\theta} \ \tilde{I}_j^x,\nonumber\\
I_j^x &=& e^{i\theta \tilde{I}_j^y} \tilde{I}_j^x e^{-i\theta \tilde{I}_j^y} = \cos{\theta} \ \tilde{I}_j^x  + \sin{\theta} \ \tilde{I}_j^z,\nonumber\\
I_j^y &=&  \tilde{I}_j^y,
\end{eqnarray}
where $\cos{\theta} = \frac{-\Delta}{\sqrt{\Delta^2 + \Omega^2 }}$ and $\sin{\theta} = \frac{\Omega}{\sqrt{\Delta^2 + \Omega^2 }}$.

Now it is easy to demonstrate that 
\begin{equation}
H = \omega_1 \tilde{I}^z_1 +  \omega_2 \tilde{I}^z_2 +  \lambda_{1,2} \left[I^z_1 I^z_2 -\frac{1}{2}(I^x_1 I^x_2 + I^y_1 I^y_2 )  \right] \approx \omega_1 \tilde{I}^z_1 +  \omega_2 \tilde{I}^z_2
\end{equation}
when $\theta = \theta_m=\arctan{( \sqrt{2} )} \approx 54.7^\circ$, i.e. equals to the magic angle~\cite{Lee65}.

In our case, and considering only the terms in Eq.~(\ref{sup:demons}) which are relevant for the demonstration, we have that the equivalent expression to Eq.~(\ref{sup:twonuclei}) is 
\begin{equation}
H = \sum_j \bigg\{ \big[ \Delta  \tilde{n}_x + \frac{\Omega_x}{2} (1 - \tilde{n}_x^2 - \tilde{n}_z) + \frac{m_s}{2} \tilde{n}_z \ \tilde{n}_x \ A_j^z \big] \ I_j^x + \big[\Delta \tilde{n}_z + \frac{\Omega_x}{2}\tilde{n}_x (1 - \tilde{n}_z)+ \frac{m_s}{2}  \ \tilde{n}^2_z \ A_j^z \big]  \ I_j^z  \bigg\}v+ e^{i \xi  \sum_j\tilde{n} \cdot \vec{I}_j t}  \tilde{H}_{nn} e^{-i \xi  \sum_j\tilde{n} \cdot \vec{I}_j t}.
\end{equation}
Now taking into account that  $\tilde{n} = (\frac{\Omega_x}{\tilde\omega}, 0,  -\frac{\omega + \omega_{\rm rf}}{\tilde\omega}) \approx (0, 0,  -1)$, which is a reasonable assumption in the parameter regime we are considering, we can simplify the above expression to find

\begin{equation}\label{generalsimplifyed}
H \approx \sum_j     \Omega_x   \ I_j^x + \sum_j   \big[- \Delta  + \frac{m_s}{2}   \ A_j^z \big]  \ I_j^z  +  \tilde{H}_{nn} = \sum_j    \Omega_x   \ I_j^x -\Delta \sum_j   \big[1  -   \frac{m_s A_j^z}{2 \Delta} \big]  \ I_j^z  +  \tilde{H}_{nn}.
\end{equation}
In this manner one can apply the same formalism to search for the magic angle condition  but including a correction of the order $\frac{ A_j^z}{2 \Delta}$  (note that $|m_s| = 1$). Hence we can define a new basis such that 
\begin{eqnarray}
I_j^z &=& e^{i\theta_j \tilde{I}_j^y} \tilde{I}_j^z e^{-i\theta_j \tilde{I}_j^y} \approx  e^{i\theta_m \tilde{I}_j^y} \tilde{I}_j^z e^{-i\theta_m \tilde{I}_j^y} (1 + O\left[\big| A_j^z /2 \Delta \big|\right] )= \cos{\theta_m} \ \tilde{I}_j^z - \sin{\theta_m} \ \tilde{I}_j^x + O\left[\big| A_j^z /2 \Delta \big|\right],\nonumber\\
I_j^x &=& e^{i\theta_j \tilde{I}_j^y} \tilde{I}_j^x e^{-i\theta_j \tilde{I}_j^y} \approx  e^{i\theta_m \tilde{I}_j^y} \tilde{I}_j^x e^{-i\theta_m \tilde{I}_j^y} (1 + O\left[\big| A_j^z /2 \Delta \big|\right] )=  \cos{\theta} \ \tilde{I}_j^x  + \sin{\theta} \ \tilde{I}_j^z + O\left[\big| A_j^z /2 \Delta \big|\right] ,\nonumber\\
I_j^y &=&  \tilde{I}_j^y,
\end{eqnarray}
In this manner we have that the Hamiltonian in Eq.~(\ref{generalsimplifyed}) is 
\begin{equation}
H \approx \sum_j     \Omega_x   \ I_j^x + \sum_j   \big[- \Delta  + \frac{m_s}{2}   \ A_j^z \big]  \ I_j^z  +    O\left[\big| A_j^z /2 \Delta \big|\right].
\end{equation}
Therefore the each nuclear-nuclear interaction is suppressed by a factor of $\big|A_j^z /2 \Delta\big|$.
\section{Numerical simulations}
\subsection{Figure 2 sample}
Figure $2$ of the main text simulate a sample consisting on an NV center and a nuclear spin cluster located at positions
\begin{eqnarray}
\vec{r}_1 &=& [-0.1262,   \ 0.8016,   \ 0.2061 ] \ \mbox{nm},\\\nonumber
\vec{r}_2 &=& [0.2524,    \ -0.5830, \   -0.3607 ] \ \mbox{nm},\\\nonumber
\vec{r}_3 &=& [0.2524,  \  -0.7287,    \ -0.4122 ] \ \mbox{nm}.
\end{eqnarray}
 Figure $2$ a) uses $B_z =0.1$ T, while in  Fig. $1$ b)  a high magnetic field of $B_{z} = 2$ T, and a value of $\Delta = 2\pi \times 100$ kHz have been employed.

\subsection{Table 1 sample}
In Table $1$ we used
\begin{eqnarray}
\vec{r}_1 &=& [0.1262,   \ 0.8016,   \  0.8245] \ \mbox{nm},\\\nonumber
\vec{r}_2 &=& [-0.6311,    \ -0.2186, \   0.6183 ] \ \mbox{nm},\\\nonumber
\vec{r}_3 &=& [-0.6311,  \  -0.2644,    \ 0.8760] \ \mbox{nm},
\end{eqnarray}
with $B_z = 2$ T  and  $\Delta = 2\pi \times 100$ kHz. 

\subsection{Fidelities under strong error conditions}
In Table~\ref{sup:table1} We show the fidelities for the sample used in the main text under strong error conditions $\Lambda = 2\pi \times 2$ MHz, and Rabi frequency error of $5 \%$.

\begin{table}[h] 
\centering
\caption{}
\label{sup:table1}
\vspace{2.0mm}
\begin{tabular}{{ |c | c | c | c | c| c|}}
 \hline
  $F_{-,+}$                        & exp$( \mp i\frac{\pi}{2} \sigma_z I_j^x)$&  exp$(\mp i\frac{\pi}{2} \sigma_z I_j^y)$ &    exp$( \mp i\frac{\pi}{2}  I_j^x)$ &  exp$( \mp i\frac{\pi}{2}  I_j^y)$\\
\hline
Spin$_1$  & $ 0.9858,  0.9861 $    &$0.9766, 0.9831$      & $0.9861, 0.9862$     &$0.9839, 0.9849$  \\
\hline
Spin$_2$    & $0.9802, 0.9802$    &$0.9684, 0.9773$      & $  0.9924, 0.9924$   & $0.9924, 0.9923$ \\
\hline
Spin$_3$  & $0.9893, 0.9852$    &$0.9738, 0.9778$    & $0.9917, 0.9915$  & $ 0.9911, 0.9913$ \\
\hline
\end{tabular}
\end{table}

\subsection{Nuclear bath contribution}
To estimate the effects of the nuclear environment we deal with $10$ different samples containing $200$ $^{13}$C nuclei, i.e. with $10$ different nuclear distributions that interfere  with the NV center and the three qubit nuclear register, and obtain the results outlined in Table~\ref{sup:table2}.  Note that because of machine restrictions we deal with instantaneous microwave pulses.
\begin{table*}[h!]
\centering
\caption{Coherence  ($L$) for different samples. $d_{\rm min}$ is the distance of the closest nucleus belonging to the nuclear bath to the NV center, while $d_{\rm max}$ is the distance of the furthest one.  The nuclei that conform the bath are randomly generated  according to the diamond  lattice available positions between the radius $d_{\rm min}$ and $d_{\rm max}$.}
\label{sup:table2}
\vspace{2.0mm}
\begin{tabular}{{ |c | c | c | c| c|}}
 \hline
          &   $L$  &   $d_{\rm min}$ (nm) & $d_{\rm max}$  (nm)\\
\hline
Sample 1  & $0.9939$   &  $1.4584$   &$4.7429$ \\
\hline
Sample 2  & $0.9949$   &  $1.5433$   &$4.8623 $ \\
\hline
Sample 3  & $0.9921$ & $1.3913$ & $4.8492$\\
\hline
Sample 4  & $0.9932$&$1.4140$ &$4.8096$\\
\hline
Sample 5  & $0.9946$&$ 1.5840$ &$4.8096$\\
\hline
Sample 6  & $0.9936$&$ 1.5840$ &$4.7319$\\
\hline
Sample 7  & $0.9938$&$ 1.5433$ &$4.7387$\\
\hline
Sample 8  & $0.9926$&$ 1.3117$ &$4.8360$\\
\hline
Sample 9  & $0.9932$&$1.6934$ &$4.7722$\\
\hline
Sample 10  & $0.9940$&$1.5225$ &$4.8492$\\
\hline
\end{tabular}
\end{table*}


\begin{thebibliography}{37}
\bibitem{Feynman82} R. P. Feynman, Int. J. Theor. Phys. {\bf 21}, 467 (1982).

\bibitem{Lloyd96}  S. Lloyd, Science {\bf 273}, 1073 (1996).

\bibitem{LeibfriedEtAl} D. Leibfried, R. Blatt, C. Monroe, and D. Wineland, Rev. Mod. Phys. {\bf 75}, 281 (2003).

\bibitem{Devoret13} M. H. Devoret and R. J. Schoelkopf, Science {\bf 339}, 1169 (2013).

\bibitem{Bloch05} I. Bloch, Nat. Phys. {\bf 1}, 23 (2005).

\bibitem{HartmannBP08} M.J. Hartmann, F.G.S.L. Brand{\~a}o and M.B. Plenio, Laser \& Photonics 
Reviews {\bf 6}, 527 (2008).

\bibitem{Obrien09} J. L. O'Brien, A. Furusawa, and J. Vu\v{c}kovi\'{c}, Nat. Phot. {\bf 3}, 687 (2009).

\bibitem{Doherty13} M. W. Doherty, N. B. Manson, P. Delaney, F. Jelezko, J. Wrachtrup and L. C. L. Hollenberg, Phys. Reports \textbf{528}, 1 (2013).

\bibitem{GruberDT+1997} A. Gruber, A. Dr{\"a}benstedt, C. Tietz, L. Fleury, J. Wrachtrup and C. von Borczyskowski,
Science {\bf 276}, 2012 (1997).

\bibitem{GaebelDP+2006} T. Gaebel, M. Domhan, I. Popa, C. Wittmann, P. Neumann, F. Jelezko,
J. R. Rabeau, N. Stavrias, A. D. Greentree, S. Prawer, J. Meijer, J. Twamley, P. R. Hemmer and
J. Wrachtrup, Nat. Phys. {\bf 2}, 408 (2006).

\bibitem{Gurudev07}  M. V. Gurudev Dutt, L. Childress, L. Jiang, E. Togan, J. Maze, F. Jelezko, A. S. Zibrov, P. R. Hemmer, and M. D. Lukin, Science {\bf 316}, 1312 (2007).

\bibitem{Neuman10} P. Neumann, J. Beck, M. Steiner, F. Rempp, H. Fedder, P. R. Hemmer, J. Wrachtrup, and F. Jelezko, Science {\bf 329}, 542 (2010).

\bibitem{Robledo11} L. Robledo, L. Childress, H. Bernien, B. Hensen, P. F. A. Alkemade, and R. Hanson, Nature {\bf 477}, 574 (2011).

\bibitem{vanderSar12} T. van der Sar, Z. H. Wang, M. S. Blok, H. Bernien, T. H. Taminiau, D. M. Toyli, D. A. Lidar,	D. D. Awschalom, R. Hanson, and V. V. Dobrovitski,  Nature {\bf 484}, 82 (2012).

\bibitem{Kolkowitz12} S. Kolkowitz, Q. P. Unterreithmeier,  S. D. Bennett, and M. D. Lukin, Phys. Rev. Lett. {\bf 109}, 137601 (2012).

\bibitem{Taminiau12} T. H. Taminiau, J. J. T. Wagenaar, T. van der Sar, F. Jelezko, V. V. Dobrovitski, and R. Hanson, Phys. Rev. Lett. {\bf 109}, 137602 (2012).


\bibitem{Liu13} G.-Q. Liu, H. C. Po, J. Du, R.-B. Liu, and X.-Y. Pan, Nat. Comm. {\bf 4}, 2254 (2013). 


\bibitem{Taminiau14} T. H. Taminiau, J. Cramer, T. van der Sar, V. V. Dobrovitski, and R. Hanson, Nature Nanotech. {\bf 9}, 171  (2014).

\bibitem{WaldherrWZ+2014} G. Waldherr, Y. Wang, S. Zaiser, M. Jamali, T. Schulte-Herbr{\"u}ggen, H. Abe, T. Ohshima, J. Isoya, J.F. Du, P. Neumann, and J. Wrachtrup, Nature {\bf 506}, 204 (2014).

\bibitem{Maurer12} P. C. Maurer, G. Kucsko, C. Latta, L. Jiang, N. Y. Yao,  S. D. Bennett, F. Pastawski, D. Hunger,  N. Chisholm, M. Markham, D. J. Twitchen, J. I. Cirac, and M. D. Lukin, Science {\bf 336}, 1283 (2012).

\bibitem{Nielsen} M. A. Nielsen and I. L. Chuang, {\it Quantum Computation and Quantum Information} (Cambridge University press, Cambridge, 2000).

\bibitem{KnillL1998} E. Knill and R. Laflamme, Phys. Rev. Lett. {\bf 81}, 5672 (1998).

\bibitem{ParkerP2000} S. Parker and M. B. Plenio, Phys. Rev. Lett. {\bf 85}, 3049 (2000).

\bibitem{Fowler12} A. G. Fowler, A. C. Whiteside, and L. C. L. Hollenberg, Phys. Rev. Lett. {\bf 108}, 180501 (2012).

\bibitem{Supplemental} Supplemental Material.

\bibitem{Casanova15} J. Casanova, Z.-Y. Wang, J. F. Haase, and M. B. Plenio, Phys. Rev. A {\bf 92}, 042304 (2015).

\bibitem{Wang15}  Z.-Y. Wang, J. F. Haase, J. Casanova, and M. B. Plenio, Phys. Rev. B {\bf 93}, 174104 (2016).

\bibitem{Lee65} M. Lee and W. I. Goldburg, Phys. Rev. {\bf 140}, A1261 (1965).

\bibitem{CaiRJ+2013} J. M. Cai, A. Retzker, F. Jelezko and M. B. Plenio, Nat. Phys. {\bf 9}, 168 (2013).

\bibitem{Michal08} C. A. Michal, S. P. Hastings, and L. H. Lee, The Journal of Chemical Physics, {\bf 128}, 052301 (2008).

\bibitem{Carr54} H. Y. Carr and E. M. Purcell, Phys. Rev. \textbf{94}, 630 (1954).

\bibitem{Meiboom58} S. Meiboom and D. Gill, Rev. Sci. Instrum. {\bf 29}, 688 (1958).

\bibitem{Maudsley86} A. A. Maudsley, J. Magn. Reson. {\bf 69}, 488 (1986).

\bibitem{Gullion90} T. Gullion, D. B. Baker, and M. S. Conradi, J. Magn. Reson. {\bf 89}, 479 (1990).

\bibitem{Loretz15} M. Loretz, J. M. Boss, T. Rosskopf, H. J. Mamin,
D. Rugar, and C. L. Degen, Phys. Rev. X \textbf{5}, 021009 (2015).

\bibitem{Faraday15} M. Orrit et al., Faraday Discuss. {\bf 184}, 275 (2015).

\bibitem{Jarmola12} A. Jarmola, V. M. Acosta, K. Jensen,  S. Chemerisov,  and D. Budker, Phys. Rev. Lett. {\bf 108}, 197601 (2012).

\bibitem{Cramer15} J. Cramer, N. Kalb, M. A. Rol, B. Hensen, M. S. Blok, M. Markham, D. J. Twitchen, R. Hanson, T. H. Taminiau, Nat. Commun. {\bf 7}, 11526 (2016).

\bibitem{Maze08} J. R. Maze, J. M. Taylor, and M. D. Lukin, Phys. Rev. B {\bf 78}, 094303  (2008). 

\bibitem{Zhao12} N. Zhao, J. Honert, B. Schmidt, M. Klas, J. Isoya, M. Markham, D. Twitchen, F. Jelezko, R.-B. Liu, H. Fedder, and J. Wrachtrup, Nat. Nanotechnology. {\bf 7} 657 (2012).

\bibitem{Cai12} J. M. Cai,  B. Naydenov, R. Pfeiffer, L. P. McGuinness, K. D. Jahnke, F. Jelezko, M. B. Plenio, and A. Retzker, New. J. Phys. {\bf 14},  113023 (2012).

\bibitem{Wang08} X. Wang, C. -S. Yu, and X. X. Yi, Phys. Lett. A {\bf 373}, 58 (2008).

\bibitem{Jordan28} P. Jordan and E. Wigner, Z. Phys. {\bf 47}, 631 (1928).

\bibitem{Casanova12} J. Casanova, A. Mezzacapo, L. Lamata, and E. Solano, Phys. Rev. Lett. {\bf 108}, 190502 (2012).

\bibitem{Lamata14} L. Lamata, A. Mezzacapo, J. Casanova, and E. Solano, EPJ Quantum Technology {\bf 1}, 9 (2014).

\bibitem{ParkerP2002} S. Parker and M. B. Plenio, J. Mod. Opt. {\bf 49}, 1325 (2002).

\bibitem{BoixoS08} S. Boixo and R.D. Somma, Phys. Rev. A {\bf 77}, 052320 (2008).

\end{thebibliography}

\begin{thebibliography}{37}
\bibitem{Casanova15} J. Casanova, Z.-Y. Wang, J. F. Haase, and M. B. Plenio, Phys. Rev. A {\bf 92}, 042304 (2015).
\bibitem{Allen} L. Allen and J. H. Eberly, {\it Optical Resonance and Two-Level Atoms}  (John Wiley \& Sons, 1975).  
\bibitem{Lee65}  M. Lee and W. I. Goldburg, Phys. Rev. {\bf 140}, A1261  (1965).
\end{thebibliography}
\end{document}